\documentclass[prd,amsmath,amssymb,aps,twocolumn]{revtex4-1}

\usepackage{subcaption}
\usepackage[dvipsnames]{xcolor}
\usepackage{graphicx}
\usepackage{comment}
\usepackage[colorlinks=true,citecolor=blue]{hyperref}
\usepackage{ulem}
\usepackage{soul}
\setstcolor{Green}
\usepackage{cancel}
\usepackage{mathrsfs}
\DeclareMathAlphabet{\mathcalligra}{T1}{calligra}{m}{n}
\usepackage[utf8]{inputenc}

\newcommand{\subR}{\textrm{\tiny{R}}}
\newcommand{\subQ}{\textrm{\tiny{Q}}}
\newcommand{\subE}{\textrm{\tiny{E}}}

\newcommand{\subB}{\textrm{\tiny{B}}}
\newcommand{\subU}{\textrm{\tiny{U}}}
\newcommand{\subHH}{\textrm{\tiny{HH}}}

\newcommand{\be}{\begin{equation} }
	\newcommand{\ee}{\end{equation}}
\newcommand{\bes}{\begin{equation*} }
	\newcommand{\ees}{\end{equation*}}
\newcommand{\bea}{\begin{eqnarray} }
	\newcommand{\eea}{\end{eqnarray}}
\newcommand{\beas}{\begin{eqnarray*} }
	\newcommand{\eeas}{\end{eqnarray*}}
\newcommand{\ba}{\begin{align} }
	\newcommand{\ea}{\end{align} }
\newcommand{\bas}{\begin{align*} }
	\newcommand{\eas}{\end{align*} }

\usepackage{caption}
\newcommand{\jar}[1]{\slshape{\color{violet}}}
\newcommand{\scri}{{\mathscr I}}
\begin{document}
	
	\title{Renormalized stress-energy tensor for scalar fields in Hartle-Hawking, Boulware and Unruh states in the Reissner-Nordström spacetime}
	\author{Julio Arrechea}
	\email{arrechea@iaa.es}
	\affiliation{Instituto de Astrof\'isica de Andaluc\'ia (IAA-CSIC), Glorieta de la Astronom\'ia, 18008 Granada, Spain}
	
	\author{Cormac Breen}
	\email{cormac.breen@tudublin.ie}
	\affiliation{School of Mathematics and Statistics, Technological University Dublin, Grangegorman, Dublin 7, Ireland}

\author{Adrian Ottewill}
	\email{adrian.ottewill@ucd.ie}
	\affiliation{School of Mathematics and Statistics, University College Dublin, Belfield, Dublin 4, Ireland}
	
\author{Peter Taylor}
	\email{peter.taylor@dcu.ie}
	\affiliation{
		Center for Astrophysics and Relativity, School of Mathematical Sciences, Dublin City University, Glasnevin, Dublin 9, Ireland}
		
	\date{\today}
	\begin{abstract}
	In this paper, we consider a quantum scalar field propagating on the Reissner-Nordström black hole spacetime. We compute the renormalized stress-energy tensor for the field in the Hartle-Hawking, Boulware and Unruh states. When the field is in the Hartle-Hawking state, we renormalize using the recently developed ``extended coordinate'' prescription. This method, which relies on Euclidean techniques, is very fast and accurate. Once, we have renormalized in the Hartle-Hawking state, we compute the stress-energy tensor in the Boulware and Unruh states by leveraging the fact that the difference between stress-energy tensors in different quantum states is already finite. We consider a range of coupling constants and masses for the field and a range of electric charge values for the black hole, including near-extreme values. Lastly, we compare these results with the analytic approximations available in the literature.
	\end{abstract}
	\maketitle
	
 \section{Introduction}
 	The renormalized expectation value of the quantum stress-energy tensor plays a crucial role in the semi-classical theory of gravity. It governs the quantum backreaction on the classical spacetime geometry via the semi-classical field equations 	
  \begin{align}\label{eq:semieqs}
		G_{ab}-\Lambda\,g_{ab}+\alpha H^{(1)}_{ab}+\beta\,H^{(2)}_{ab}=8\pi\,\langle\hat{T}_{ab}\rangle_{\subR},
	\end{align}
where $g_{ab}$ is the metric of spacetime, $G_{ab}$ is the Einstein tensor and $\langle\hat{T}_{ab}\rangle_{\subR}$ is the renormalized expectation value of the stress-energy tensor of a quantum field in some quantum state. The tensors $H^{(1)}_{ab}$, $H^{(2)}_{ab}$ are geometrical terms that are quadratic in the curvature and arise through the point-splitting regularization process that yields $\langle\hat{T}_{ab}\rangle_{\subR}$. This regularization process corresponds to an infinite renormalization of the constants $\Lambda$, $\alpha$ and $\beta$. 

The calculation of renormalized expectation values of the stress-energy tensor (RSET) for a particular quantum state is a technically challenging endeavour. For black hole spacetimes, there are three main approaches to this calculation, the most established being the Candelas-Howard approach \cite{CandelasHoward:1984} and its extensions (see for example \cite{AHS1995, TaylorOttewill2010, BreenOttewill2012}). Recently two new, more efficient methods for the calculation of the RSET have been developed, the first being the ``pragmatic mode-sum prescription" \cite{LeviOri:2015, LeviOri:2016}. The method has proven indeed to be pragmatic, both in its efficiency and its broader applicability. The second recent development in methods to compute the RSET in black hole spacetimes is known as the ``extended coordinate method" \cite{taylorbreen:2016,taylorbreen:2017, taylorbreenottewill:2021}, which is extremely efficient and applicable to arbitrary field parameters and arbitrary spacetime dimensions. 

Of the three standard quantum states considered in spherically symmetric black hole space-times, the Hartle-Hawking state \cite{HartleHawking:1976, Jacobson:1994} has received the most attention (see for example \cite{CandelasHoward:1984, Howard:1984,JensenOttewill:1989, AHS1995, Winstanley:2008, BreenOttewill2012, BreenOttewill:2010, TaylorOttewill2010}). This is mainly due to the fact that this state can be defined on the Euclidean metric and there are some convenient simplifications that occur in this Euclidean setting, for example, the frequency spectrum is discrete. When it comes to the other two quantum states of interest, the Boulware and Unruh states \cite{Boulware:1975,Unruh:1976}, the literature is more sparse. Anderson, Hiscock and Samuel developed a method for the calculation of the RSET of a scalar field with arbitrary mass and coupling in the Boulware state in a general spherically symmetric (Euclidean) metric \cite{AHS1995}. Jensen, Mc Laughlin and Ottewill calculated the RSET for a massless, conformally coupled field in the Unruh and Boulware states in the Schwarzschild spacetime \cite{Jensen:1991, Jensen:1992}. More recently, the pragmatic mode-sum prescription has been applied to the calculation of the RSET for the Boulware and Unruh states for a massless minimally coupled field in the Schwarzschild, Reissner-Nordström and Kerr spacetimes \cite{LeviOri:2015,LeviOri:2016, LeviOri:2016b}. Anderson, Siahmazgi, Clark and Fabbri also applied the pragmatic mode-sum prescription to the calculation of the RSET for a massless minimally coupled field in a spacetime where a black hole forms from the collapse of a null shell \cite{Anderson:2020}. The relative lack of RSET results for the Unruh state and the fact that there does not appear to be any results whatsoever in the literature for a massive field in this state is surprising, given that the Unruh state is considered to be the one of most physical interest, that is, the state that models the late-time evolution of the collapse of a spherical body to a black hole \cite{Unruh:1976}.

The extended coordinate renormalization method \cite{taylorbreen:2016,taylorbreen:2017, taylorbreenottewill:2021} mentioned above was developed using Euclidean techniques and is applicable only to a quantum field in the Hartle-Hawking state. While it is possible, by considering a field at zero temperature in the Euclidean metric, to extend the method to the Boulware vacuum, it is more efficient to adopt a state subtraction approach using the Hartle-Hawking as our reference state. In the state subtraction approach one leverages the fact that the difference between stress tensors in different quantum states does not require renormalization, therefore by calculating the RSET in the Hartle-Hawking state we may then obtain the RSET in the Unruh and Boulware states without recourse to renormalization.  We note that for the case of a spacetime where we can not define a Hartle-Hawking vacuum to use as our reference state (such as in Kerr black holes~\cite{OttewillWinstanley2000}), we must then consider the direct calculation for the Boulware state through the zero temperature extended coordinate method mentioned above. We hope to present results on this in the near future.

In this paper we employ the extended coordinate method to compute the RSET for a scalar field in the Hartle Hawking state propagating in a general spherically symmetric black hole spacetime. We assume the field has arbitrary mass and coupling to the background curvature. We then show how to compute the RSET in the Boulware and Unruh states by a state subtraction scheme, using the RSET in the Hartle-Hawking state as a reference state. We then apply these results to the particular case of the Reissner-Nordström spacetime and calculate the RSET in all three quantum states for a range of quantum field masses and coupling constants. We also consider a range of electric charge values for the black hole and probe the near extreme case. Finally we compare the exact RSET results with various analytic approximations and judge their reliability.

\section{Renormalization Prescription in the Hartle-Hawking State}
 \label{sec:HH}
In this section, we will briefly outline the extended coordinate approach to calculating the RSET for a quantum scalar field in the Hartle-Hawking quantum state, propagating on a static, spherically symmetric black hole spacetime.  It is convenient to construct the Hartle-Hawking state on the Euclideanized line element
\begin{align}
	\label{eq:metric}	ds^{2}=f(r)d\tau^{2}+dr^{2}/f(r)+r^{2}( d\theta^2 + \sin^2 \theta d \phi^2),	\end{align}
where in order to avoid a conical singularity on the event horizon $r=r_{+}$, it is necessary to impose the periodicity \mbox{$\tau=\tau+2\pi/\kappa_+$} on the Euclidean time, where 
\begin{equation}
    \kappa_+=\frac{1}{2}f'(r)\big|_{r=r_{+}},
\end{equation}
is the horizon surface gravity. Imposing this periodicity discretizes the frequency spectrum of the field modes which now satisfy an elliptic equation
\begin{align}\label{eq:kg}
	(\Box_{\subE}-\mu^2-\xi\,R)\phi=0,
\end{align}
where $\Box_{\subE}$ is the d'Alembertian operator with respect to the Euclidean metric, $\mu$ is the field mass, $R$ is the Ricci curvature scalar of the background spacetime and $\xi$ is the coupling strength between the field and the background geometry. The corresponding Euclidean Green function has the following mode-sum representation
(with $r=r'$ for simplicity) on this black hole spacetime,
\begin{align}
	\label{eq:Gmodesum}
	&G(x,x')=\frac{1}{8 \pi^2}\sum_{l=0}^{\infty}(2l+1)P_l(\cos\gamma)\sum_{n=-\infty}^{\infty}e^{in\kappa_+\Delta\tau}g_{nl}(r),
\end{align}
where $\Delta x\equiv x'-x\sim\mathcal{O}(\epsilon)$ is the coordinate separation, $\gamma$ is the geodesic distance on the 2-sphere and $P_{l}(z)$ is the Legendre polynomial of the first kind. We have denoted by $g_{nl}(r):=\kappa_+\,p_{nl}(r)\,q_{nl}(r)/N_{nl}$ the one-dimensional radial Green function evaluated at the same spacetime point $r$. The radial modes $p_{nl}(r)$, $q_{nl}(r)$ are solutions of the homogeneous radial equation:
\begin{align}
\label{eq:Euclideanradialeqn}
	\bigg[\frac{d}{dr}\Big(r^2f(r)\frac{d}{dr}\Big)-r^{2}\left(\frac{n^{2}\kappa_+^{2}}{f(r)}+(\mu^{2}+\xi\,R)\right)\nonumber\\
	-l(l+1)\bigg]\Phi_{nl}(r)=0,
\end{align}
where $p_{nl}(r)$ and $q_{nl}(r)$ are regular on the horizon and the outer boundary (usually spatial infinity), respectively. The normalization constant is given by
\begin{align}
	N_{nl}=-r^{2}f(r)\,\mathcal{W}\{p_{nl}(r),q_{nl}(r)\},
\end{align}
where $\mathcal{W}\{p,q\}$ denotes the Wronskian of the two solutions.

In the coincidence limit $\Delta x\to 0$ (i.e. $\gamma\to 0$ and $\Delta\tau\to 0$), the mode sum (\ref{eq:Gmodesum}) diverges. To renormalize this mode sum, we must find a way to express the locally-constructed Hadamard parametrix $K(x,x')$ as a mode sum and subtract mode-by-mode. In \cite{taylorbreen:2016,taylorbreen:2017} a mode sum expression for the Hadamard parametrix was derived by first introducing the so-called extended coordinates:
\begin{align}  
	\varpi^{2}=
 &
 \frac{2}{\kappa_+^{2}}(1-\cos \kappa_+\Delta\tau),\nonumber\\ s^{2}=
 &
 f(r)\,\varpi^{2}+2 r^{2}(1-\cos\gamma).
\end{align}
For simplicity, the separation in the radial direction, $\Delta r$, is set to zero but it is important to the development that the separation in the other directions is maintained. Expressing the Hadamard parametrix in terms of these extended coordinates permits  
its decomposition in terms of Fourier frequency modes and multipole moments where, remarkably, the coefficients in this decomposition are expressible in closed form for any static spherically-symmetric spacetime in arbitrary dimensions. In four dimensions, the result is
\begin{align}
	\label{eq:HadamardExp}
	K(x,x')=&\frac{1}{8\pi^{2}}\sum_{l=0}^{\infty}(2l+1)P_{l}(\cos\gamma)\sum_{n=-\infty}^{\infty}e^{in\kappa_+\Delta\tau}k_{nl}(r)\nonumber\\
	&+\frac{1}{8 \pi^2}\Big\{\mathcal{D}^{(-)}_{11}(r) + \left(\mathcal{T}^{(p)}_{10}+	\mathcal{D}^{(-)}_{22}(r)\right) s^2\nonumber\\
	&+\left(\mathcal{T}^{(p)}_{11}+\mathcal{D}^{(-)}_{21}(r)\right)\varpi^2 \Big\}+ O(\epsilon^{2m} \log \epsilon),
\end{align}
where
\begin{align}
	\label{eq:RegParam}
	k_{nl}(r)=& \sum_{i=0}^{m}\sum_{j=0}^{i}\mathcal{D}^{(+)}_{ij}(r)\Psi^{(+)}_{nl}(i,j|r)\nonumber\\
 &
 +\sum_{i=0}^{m-1}\sum_{j=0}^{i}\mathcal{T}^{(l)}_{ij} \chi_{nl}(i,j|r)\nonumber\\
	&+\sum_{i=1}^{m-1}\sum_{j=0}^{i-1}\mathcal{T}^{(r)}_{ij} \Psi^{(+)}_{nl}(i+1,j|r).
\end{align}
Here $m$ denotes the order of the expansion, the coefficients $\mathcal{D}^{(\pm)}_{ij}(r)$, $\mathcal{T}^{(l)}_{ij}(r)$, $\mathcal{T}^{(p)}_{ij}(r)$ and $\mathcal{T}^{(r)}_{ij}(r)$ arise in the expansion of the Hadamard parametrix $K(x,x')$ in the extended coordinates $s$ and $\varpi$ while the terms $\Psi_{nl}^{(+)}(i,j|r)$ and $\chi_{nl}(i,j|r)$ are the so-called regularization parameters that arise in  expressing $K(x,x')$ as a mode sum. The well-known renormalization ambiguity is expressed as an arbitrary lengthscale in the regularization parameters $\chi_{nl}(i,j|r)$. We find that all the regularization parameters are obtainable in closed form in terms of complicated combinations of special functions. Explicit expressions for each of these are given in \cite{taylorbreenottewill:2021}.

To apply the extended coordinate method to the calculation of the RSET, we first introduce some notation:
\begin{align}
	&W(x,x')\equiv G(x,x')-K(x,x')\nonumber\\
	&\langle \hat{\phi}^{2}\rangle_{\subR}\equiv \left[W(x,x')\right]\equiv w(r)\nonumber\\
	&w_{ab}(x)\equiv\left[W(x,x')_{;a'b'}\right] = \left[W(x,x')_{;ab}\right],
\end{align}
were we have adopted square brackets $[\cdot]$ to denote the coincidence limit $x'\to x$. Then the RSET may be written in the form \cite{taylorbreenottewill:2021}:
\begin{align}
\label{eq:RSET}
	\langle \hat{T}_\xi^{a}{}_{b}\rangle_{\subR} & = - w^{a}{}_{b}   -   (\xi-\tfrac{1}{2})\langle\hat{\phi}^2\rangle_{\subR}{}^{;a}{}_{;b}+ (\xi-\tfrac{1}{4}) \square \langle\hat{\phi}^2\rangle_{\subR} \delta^{a}{}_{b}  \nonumber\\
	&\qquad +\xi R^{a}{}_{b} \langle\hat{\phi}^2\rangle_{\subR} -  \frac{1}{8\pi^2} v_1 \delta^{a}{}_{b}.
\end{align}
where:
\begin{align}
	v_1 &= \tfrac{1}{720}R_{pqrs}R^{pqrs}- \tfrac{1}{720}R_{pq}R^{pq}- \tfrac{1}{24}(\xi-\tfrac{1}{5})\square R\nonumber\\
	&\quad  + \tfrac{1}{8}(\xi-\tfrac{1}{6})^2R^2 + \tfrac{1}{4}\mu^2(\xi-\tfrac{1}{6}) R +\tfrac{1}{8}\mu^4.
\end{align}
We note that since $\langle\hat\phi^2\rangle_{\subR}$ is a function of $r$ only, once it has been calculated numerically to high accuracy on a suitaby dense grid, derivatives of $\langle\hat\phi^2\rangle_{\subR}$ are easily and accurately obtained by differentiating an interpolation function for $\langle\hat\phi^2\rangle_{\subR}$. Considering next the components of $ w^{a}{}_{b}$, we have, by virtue of the wave equation satsified by $W(x,x')$, that~\cite{BrownOttewill:1986}: 
 \begin{align}
	w^{r}{}_{r}  = -w^{\tau}{}_{\tau}-w^{\theta}{}_{\theta}-w^{\phi}{}_{\phi} -  \xi R w  -\mu^2 w  -\frac{3}{4\pi^2}v_1.
\end{align}
For the remaining non-zero components, $w^{\tau}{}_{\tau}$ and \mbox{$w^{\theta}{}_{\theta}=w^{\phi}{}_{\phi}$}, it is advantageous to express these in terms of mixed derivatives at $x$ and $x'$ of $W(x,x')$ and derivatives of $w(r) =\langle\hat\phi^2\rangle_{\subR}$, using Synge's Rule \cite{taylorbreenottewill:2021}:
 \begin{align}
	\left[W(x,x')_{;a'b}\right] =  \tfrac{1}{2} w_{;ab}(x) - w_{ab}(x).
\end{align}  
The required mixed time derivatives and mixed angular derivatives may then in turn be expressed as mode sums, given by  \cite{taylorbreenottewill:2021}:
\begin{align}
	&[g^{\tau \tau'}W_{,\tau\tau'}]=-\frac{1}{4\pi^{2}}\sum_{l=0}^{\infty}(2l+1)\sum_{n=1}^{\infty}\frac{n^{2}\kappa_+^{2}}{f(r)}\mathsf{g}_{nl}(r)\nonumber\\
	&\qquad-\frac{1}{4\pi^{2}}\left\{\mathcal{T}_{10}^{(p)}+\mathcal{D}_{22}^{(-)}+\frac{1}{f(r)}(\mathcal{T}_{11}^{(p)}+\mathcal{D}_{21}^{(-)})\right\},\\
	&[g^{\phi \phi'}W_{,\phi\phi'}]=\frac{1}{16\pi^{2}r^{2}}\sum_{l=0}^{\infty}(2l+1)l(l+1)\sum_{n=0}^{\infty}(2-\delta^{n}_{0})\mathsf{g}_{nl}(r)\nonumber\\
	&\qquad+\frac{1}{4\pi^{2}}\left\{\mathcal{T}_{10}^{(p)}+\mathcal{D}_{22}^{(-)}\right\},
	\end{align}
where $\mathsf{g}_{nl}(r)\equiv g_{nl}(r)-k_{nl}(r)$. Therefore, the application of the extended coordinates method reduces the calculation of the RSET to that of three mode sums, those just presented along with
\begin{align}
	&\langle \hat{\phi}^{2}\rangle_{\subR}=\frac{1}{8\pi^{2}}\sum_{l=0}^{\infty}(2l+1)\sum_{n=0}^{\infty}(2-\delta^{n}_{0})\mathsf{g}_{nl}-\frac{\mathcal{D}_{11}^{(-)}(r)}{8\pi^{2}}.
\end{align}
The modes $\mathsf{g}_{nl}(r)$ converge like $O(l^{-2m-3})$ for large $l$, fixed $n$ and $O(n^{-2m-3})$ for large $n$, fixed $l$, where we remind the reader that $m$ is the order of the expansion of the singular field. We therefore have precise control over the convergence of the mode sums by choosing the order of the expansion. By choosing a sufficiently high order Hadamard parametrix, very high accuracy in the RSET calculation can be achieved by truncating the sums at a modest number of $l$ and $n$ modes. 

 \section{RSET in the Boulware and Unruh States}
 \label{sec:B&U}
In this section, we will outline our strategy for computing the RSET in the Boulware and the Unruh states. There are two approaches one might consider. The first is to adapt the extended coordinate approach to decompose the Hadamard parametrix for the Lorentzian spacetime and then apply the Lorentzian equivalent of the mode sum renormalization prescription described in the previous section. While taking this direct approach has some advantages, there are subtle difficulties in dealing with the Hadamard distribution on the Lorentzian sector. The other approach, which is the one we adopt here, is to use the fact that the differences between states does not require renormalization. This is because the Hadamard singularity structure is agnostic to the quantum state. So even though the Euclideanization trick employed herein is relevant only to the Hartle-Hawking state, we can still use this state as a reference to compute the RSET in other quantum states of interest.

 To be more explicit, let us start by writing out the Wightman Green function in the Boulware, Hartle-Hawking and Unruh states in terms of their Lorentzian modes. Mode solutions to the Klein-Gordon equation for a massive scalar field have the form
 \begin{align}\label{eq:harmsph}
     u_{\omega l m}=\frac{1}{\sqrt{4 \pi \omega}}e^{-i\omega t}Y_{lm}(\theta,\phi)\Phi_{\omega l}(r),
 \end{align}
 where $Y_{lm}(\theta,\phi)$ are the spherical harmonics and $\Phi_{\omega l}(r)$ is a solution of the radial equation
 \begin{align}\label{eq:radialwave}
    \left[ \frac{d}{dr}\left(r^{2}f(r)\frac{d}{dr}\right)+\frac{\omega^{2}r^{2}}{f(r)}-l(l+1)-\mu^{2}r^{2}\right]\Phi_{\omega l}(r)=0.
 \end{align}
 It is helpful to recast this equation in Schrodinger form by writing $\Phi_{\omega l}(r)=\psi_{\omega l}(r)/r$, where $\psi_{\omega l}(r)$ satisfies
 \begin{align}
 \label{eq:Schrodinger}
 \left\{\frac{d^{2}}{dr_{*}^{2}}+V_{l}(r)\right\}\psi_{\omega l}(r)=0,
 \end{align}
 where $dr_{*}/dr=1/f$ is the radial tortoise coordinate and the potential is 
\begin{align}
     V_{l}(r)=\omega^{2}-f(r)\left(\mu^{2}+\frac{l(l+1)}{r^{2}}+\frac{f'(r)}{r}\right).
\end{align}
From here on, we will work in asymptotically flat spacetimes, so spacetimes with a non-zero cosmological constant are excluded from the analysis. Importantly, this potential asymptotes to different values at the horizon and infinity for non-zero field mass. We have $V_{l}\to \omega^{2}$ as $r\to r_{+}$ while $V_{l}(r)\to \tilde{\omega}^{2}\equiv \omega^{2}-\mu^{2}$ as $r\to\infty$. Hence the solutions to Eq.~(\ref{eq:Schrodinger}) have the asymptotic forms $\psi_{\omega l}\sim e^{\pm i \omega r_{*}}$ as $r\to r_{+}$ and $\psi_{\omega l}\sim e^{\pm i \tilde{\omega}r_{*}}$ as $r\to\infty$. On the exterior, we take as a linearly independent basis the solutions with the following boundary conditions. We label the solution $\psi_{\omega l}^{\textrm{in}}(r)$ as the one with boundary condition $e^{-i\tilde{\omega}r_{*}}$ on past null-infinity and which vanishes on the past event horizon. Note that for $|\omega|<\mu$, this would be an exponentially growing mode and the solutions would not be square integrable. Hence for the ``in''-mode, we must have the restriction $\omega>\mu$. This wave which originates at past null infinity partly reflects back to future null infinity and partly transmits to the future event horizon. In terms of the boundary conditions only on the radial function, this amounts to
 \begin{align}
 \label{eq:LorentzianBCIn}
\psi^{\textrm{in}}_{\omega l}(r)&=\begin{cases}\displaystyle{B_{\omega l}^{\textrm{in}}e^{-i\omega\,r_{*}}},\qquad & r\to r_{+},\\ \\
\displaystyle{e^{-i\tilde{\omega}r_{*}}+A_{\omega l}^{\textrm{in}}e^{i\tilde{\omega}r_{*}}}, & r\to \infty, \end{cases}
\end{align}
where $B_{\omega l}^{\textrm{in}}$ and $A_{\omega l}^{\textrm{in}}$ are respectively the dimensionsless transmission and reflection coefficients for the ``in'' mode. For the other independent solution, we label $\psi_{\omega l}^{\textrm{up}}(r)$ as the one with asymptotic form $e^{i\omega r_{*}}$ on the past event horizon and which vanishes on past null infinity. This solution represents a wave propagating out of the event horizon and being partly scattered back to the future event horizon and partly transmitted to future null infinity. Then we have the asymptotic forms
\begin{align}
\label{eq:LorentzianBCUp}
\psi^{\textrm{up}}_{\omega l}(r)&=\begin{cases}
    \displaystyle{e^{i\omega r_{*}}+A_{\omega l}^{\textrm{up}}e^{-i\omega r_{*}}},\qquad & r\to r_{+}\\ \\
    \displaystyle{B_{\omega l}^{\textrm{up}} e^{i\tilde{\omega} r_{*}}}, & r\to\infty,
\end{cases}
\end{align}
where $B_{\omega l}^{\textrm{up}}$ and $A_{\omega l}^{\textrm{up}}$ are again transmission and reflection coefficients, respectively. Note that for the ``up'' modes with $|\omega|<\mu$, the solution is oscillatory near the horizon but exponentially damped at infinity. Hence these modes are square integrable and must be included in the two-point function. The ``up'' modes with $|\omega|<\mu$ are the bound-state modes \cite{Boulware:1975}.

Taking as the basis modes to the wave operator
\begin{align}
\label{eq:normalmodes}
    u_{\omega l m}^{\textrm{in}}(x)=\frac{1}{\sqrt{4\pi\,\tilde{\omega}}}e^{-i\omega t}Y_{lm}(\theta,\phi)\Phi_{\omega l}^{\textrm{in}}(r)\nonumber\\
    u_{\omega l m}^{\textrm{up}}(x)=\frac{1}{\sqrt{4\pi\omega}}e^{-i\omega t}Y_{lm}(\theta,\phi)\Phi_{\omega l}^{\textrm{up}}(r),
\end{align}
then the normalization conditions
\begin{align}
    \langle u_{J}^{\textrm{in}}, u_{J'}^{\textrm{in}}\rangle=\langle u_{J}^{\textrm{up}}, u_{J'}^{\textrm{up}}\rangle=\delta_{J J'}
\end{align}
imply
\begin{align}
\label{eq:transmission1}
    |A^{\textrm{in}}_{\omega l}|^{2}+|B^{\textrm{in}}_{\omega l}|^{2}=1,\qquad  |A^{\textrm{up}}_{\omega l}|^{2}+|B^{\textrm{up}}_{\omega l}|^{2}=1,
\end{align}
where the inner product is defined by the following integral on an arbitrary Cauchy surface $\Sigma$ with unit future-directed normal $n^{a}$,
\begin{align}
    \langle u_{J}, u_{J'}\rangle=-i\int_{\Sigma}\left(u_{J}\nabla_{a}u_{J'}^{*}-u_{J'}^{*}\nabla_{a}u_{J}\right)n^{a}d\Sigma.
\end{align}
 In deriving the normalization conditions, we move the integral over $\Sigma$ to an integral over the past event horizon $\mathcal{H}^{-}$ plus an integral over past null infinity $\scri^{-}$, making use of the asymptotic forms (\ref{eq:LorentzianBCIn})-(\ref{eq:LorentzianBCUp}). We note the different normalizations for the ``in'' modes and ``up'' modes in (\ref{eq:normalmodes}). Moreover, the constancy of the Wronskian of linearly independent solutions of the radial equation implies
 \begin{align}
 \label{eq:transmission2}
     \omega \,B_{\omega l}^{\textrm{in}}=\tilde{\omega}\,\,B_{\omega l}^{\textrm{up}}.
\end{align}
The details of how to numerically compute the normalized modes $\{\Phi^{\textrm{in}}_{\omega l}(r), \Phi^{\textrm{up}}_{\omega l}(r)\}$ are briefly discussed in the next section.

Putting these details together, and performing the trivial $m$-sum in the spherical harmonics, we get the following representation of the two-point function in terms of the normalized modes for the field in the Boulware state \cite{Boulware:1975},
\begin{widetext}
\begin{align}
    G_{\subB}(x,x')
    =\frac{1}{8\pi}\sum_{l=0}^{\infty}(2l+1)P_{l}(\cos\gamma)\Bigg[\int_{\mu}^{\infty}d\omega\frac{1}{2\pi\,\tilde{\omega}}e^{-i\omega\,\Delta t}\Phi_{\omega l}^{\textrm{in}}(r)\,\Phi_{\omega l}^{\textrm{in}}{}^{\dagger}(r')+\int_{0}^{\infty}d\omega\frac{1}{2\pi\,\omega}e^{-i\omega\,\Delta t}\Phi_{\omega l}^{\textrm{up}}(r)\,\Phi_{\omega l}^{\textrm{up}}{}^{\dagger}(r')\Bigg].
\end{align}
For the field in the Hartle-Hawking state \cite{HartleHawking:1976} there is a thermal factor $\coth(\omega\,\pi/ \kappa_{+})$ in each of the ``in'' modes and ``up'' modes. We obtain,
\begin{align}
    G_{\subHH}(x,x')=\frac{1}{8\pi}\sum_{l=0}^{\infty}(2l+1)P_{l}(\cos\gamma)\Bigg[\int_{\mu}^{\infty}d\omega\frac{1}{2\pi\,\tilde{\omega}}e^{-i\omega\,\Delta t}\coth(\omega\,\pi/\kappa_+)\Phi_{\omega l}^{\textrm{in}}(r)\,\Phi_{\omega l}^{\textrm{in}}{}^{\dagger}(r')\nonumber\\
    +\int_{0}^{\infty}d\omega\frac{1}{2\pi\,\omega}e^{-i\omega\,\Delta t}\coth(\omega\,\pi/\kappa_+)\Phi_{\omega l}^{\textrm{up}}(r)\,\Phi_{\omega l}^{\textrm{up}}{}^{\dagger}(r')\Bigg].
\end{align}
Finally, for the field in the Unruh state \cite{Unruh:1976}, we pick up a thermal factor on the ``up'' modes but not the ``in'' modes:
\begin{align}
    G_{\subU}(x,x')=\frac{1}{8\pi}\sum_{l=0}^{\infty}(2l+1)P_{l}(\cos\gamma)\Bigg[\int_{\mu}^{\infty}d\omega\frac{1}{2\pi\,\tilde{\omega}}e^{-i\omega\,\Delta t}\Phi_{\omega l}^{\textrm{in}}(r)\,\Phi_{\omega l}^{\textrm{in}}{}^{\dagger}(r')\nonumber\\
    +\int_{0}^{\infty}d\omega\frac{1}{2\pi\,\omega}e^{-i\omega\,\Delta t}\coth(\omega\,\pi/\kappa_+)\Phi_{\omega l}^{\textrm{up}}(r)\,\Phi_{\omega l}^{\textrm{up}}{}^{\dagger}(r')\Bigg].
\end{align}
\end{widetext}

 Now if we consider a quantum scalar field in any reference Hadamard state $|R\rangle$, then the RSET in any other Hadamard state $|Q\rangle$ is related to the RSET in our reference state by
 \begin{align}
 \langle \hat{T}^{a}{}_{b}\rangle_{\subQ}=\langle\hat{T}^{a}{}_{b}\rangle_{\subR}-[\delta G_{\subQ}{}^{;a}{}_{b}]-(\xi-\tfrac{1}{2})\left[\delta G_{\subQ}\right]^{;a}{}_{b}\nonumber\\
 +(\xi-\tfrac{1}{4})\delta^{a}{}_{b}\Box\left[\delta G_{\subQ}\right]+\xi\,R^{a}{}_{b}\left[\delta G_{\subQ}\right],
\end{align}
where
\begin{align}
\delta G_{\subQ}=G_{\subQ}(x,x')-G_{\subR}(x,x')
\end{align}
is the difference between our two-point function in our state $|Q\rangle$ and our reference state. The explicit dependence on the field mass, i.e., the $\mu^{2}$ term appearing in the expression for the stress-energy tensor~\eqref{eq:RSET}, canceled on application of the wave equation to $\delta G_{\subQ}$; though obviously the above expression depends on $\mu$ implicitly through $\langle \hat{T}^{a}{}_{b}\rangle_{\subR}$ and through $\delta G_{\subQ}$ itself. The salient point is that $\delta G_{\subQ}$ is a smooth homogeneous solution of the wave equation, at least on any region of the spacetime where both states satisfy the Hadamard condition.

In the current context, the reference state is the Hartle-Hawking state for which we can leverage Euclidean techniques to compute $\langle \hat{T}^{a}{}_{b}\rangle_{\subHH}$, then in the Boulware and Unruh states, we have
\begin{widetext}
\begin{align}
\delta G_{\subB}&=\frac{1}{4\pi}\sum_{l=0}^{\infty}(2l+1)P_{l}(\cos\gamma)\Bigg[\int_{\mu}^{\infty}\frac{d\omega}{2\pi\tilde{\omega}}\frac{e^{-i\omega\,\Delta t}}{(1-e^{2\pi\,\omega/\kappa_{+}})}
\Phi_{\omega l}^{\textrm{in}}(r)\Phi_{\omega l}^{\textrm{in}}{}^{\dagger}(r')+\int_{0}^{\infty}\frac{d\omega}{2\pi\,\omega}\frac{e^{-i\omega\,\Delta t}}{(1-e^{2\pi\,\omega/\kappa_{+}})}
 \Phi_{\omega l}^{\textrm{up}}(r)\Phi_{\omega l}^{\textrm{up}}{}^{\dagger}(r')\Bigg],\\
\delta G_{\subU}&=\frac{1}{4\pi}\sum_{l=0}^{\infty}(2l+1)P_{l}(\cos\gamma)\Bigg[\int_{\mu}^{\infty}\frac{d\omega}{2\pi\tilde{\omega}}\frac{e^{-i\omega\,\Delta t}}{(1-e^{2\pi\,\omega/\kappa_{+}})}
\Phi_{\omega l}^{\textrm{in}}(r)\Phi_{\omega l}^{\textrm{in}}{}^{\dagger}(r')\Bigg].
\end{align}
\end{widetext}
As we can see from the exponential thermal factor in each of these expressions, they are exponentially convergent in $\omega$ and reasonably straightforward to compute.

For convenience, we give the explicit mode-sum expressions for the RSET in the Boulware and Unruh states in terms of the RSET in the Hartle-Hawking state. For the field in the Boulware state, we have
\begin{widetext}
\begin{align}\label{eq:RSETBoulware}
\langle \hat{T}^{t}{}_{t}\rangle_{\subB}&=\langle \hat{T}^{t}{}_{t}\rangle_{\subHH}+\mathcal{\tilde{J}}\left\{V_{t}\left[\Phi^{\text {in}}_{\omega\l}(r)\right]\right\}+\mathcal{J}\left\{V_{t}\left[\Phi^{\text {up}}_{\omega\l}(r)\right]\right\},\nonumber\\
\langle \hat{T}^{\phi}{}_{\phi}\rangle_{\subB}&=\langle \hat{T}^{\phi}{}_{\phi}\rangle_{\subHH}+\mathcal{\tilde{J}}\left\{V_{\phi}\left[\Phi^{\text {in}}_{\omega\l}(r)\right]\right\}+\mathcal{J}\left\{V_{\phi}\left[\Phi^{\text {up}}_{\omega\l}(r)\right]\right\},\nonumber\\
\langle \hat{T}^{r}{}_{r}\rangle_{\subB}&=\langle \hat{T}^{r}{}_{r}\rangle_{\subHH}+\mathcal{\tilde{J}}\left\{V_{r}\left[\Phi^{\text {in}}_{\omega\l}(r)\right]\right\}+\mathcal{J}\left\{V_{r}\left[\Phi^{\text {up}}_{\omega\l}(r)\right]\right\}
\end{align}
with
    \begin{align}
    \mathcal{\tilde{J}}\left\{V\left[\Phi\right]\right\}=
    &
    \frac{1}{4\pi}\sum_{l=0}^{\infty}(2l+1)\int_{\mu}^{\infty}\frac{d\omega}{2\pi\tilde{\omega}(1-e^{2\pi\,\omega/\kappa_{+}})}V\left[\Phi\right],\nonumber\\
    \mathcal{J}\left\{V\left[\Phi\right]\right\}=
    &
    \frac{1}{4\pi}\sum_{l=0}^{\infty}(2l+1)\int_{0}^{\infty}\frac{d\omega}{2\pi\omega(1-e^{2\pi\,\omega/\kappa_{+}})}V\left[\Phi\right],
    \end{align}
and
    \begin{align}
    V_{t}\left[\Phi(r)\right]=
    &
    (2\xi-\tfrac{1}{2})f\,|\Phi'(r)|^{2}-\xi\,f'\,\Phi'(r)\Phi^{*}(r)+\left[\frac{(2\xi-\tfrac{1}{2})l(l+1)}{r^{2}}-\frac{(2\xi+\tfrac{1}{2})\omega^{2}}{f}+\xi\,R^{t}{}_{t}+\left(2\xi-\tfrac{1}{2}\right)\mu^{2}\right]|\Phi(r)|^{2},\nonumber\\
    V_{\phi}\left[\Phi(r)\right]=
    &
    (2\xi-\tfrac{1}{2})f|\Phi'(r)|^{2}-\frac{2\xi\,f}{r}\,\Phi'(r)\Phi^{*}(r)+\left[\frac{2\xi\,l(l+1)}{r^{2}}-\frac{(2\xi-\tfrac{1}{2})\omega^{2}}{f}+\xi\,R^{\phi}{}_{\phi}+\left(2\xi-\tfrac{1}{2}\right)\mu^{2}\right]|\Phi(r)|^{2},\nonumber\\
    V_{r}\left[\Phi(r)\right]=
    &
    \tfrac{1}{2}f|\Phi'(r)|^{2}+2\xi\left(\frac{2 \,f}{r}+\frac{f'}{2}\right)\,\Phi'(r)\Phi^{*}(r)+\left[-\frac{l(l+1)}{2\,r^{2}}+\frac{\omega^{2}}{2\,f}+\xi\,R^{r}{}_{r}-\tfrac{1}{2}\mu^{2}\right]|\Phi(r)|^{2},
    \end{align}
\end{widetext}
where, in order to arrive at these expressions, we have made use of Synge's Rule and the radial wave equation to avoid computing second derivatives of our radial modes at the same spacetime point.

To obtain the analogous expressions for the quantum field in the Unruh state, we simply omit the contributions from the ``up'' modes in each component in~\eqref{eq:RSETBoulware}. The RSET in the Unruh state also has an off-diagonal component
\begin{align}
\label{eq:Unruhflux}
    \langle \hat{T}_{t}{}^{r}\rangle_{\subU}=\frac{1}{4\pi\,r^{2}}\int_{\mu}^{\infty}\frac{d\omega}{2 \pi\,\tilde{\omega}}\frac{\omega^{2}}{(1-e^{2\pi\,\omega/\kappa_{+}})}\sum_{l=0}^{\infty}(2l+1)|B^{\textrm{in}}_{\omega l}|^{2},
\end{align}
which gives the outgoing flux of Hawking radiation.

It is straightforward to verify that in each state $|Q\rangle$, one obtains for the trace of the conformally-coupled stress-energy tensor
\begin{align}
    \langle \hat{T}^{a}{}_{a}\rangle_{\subQ}\Big|_{\xi=1/6}=\frac{1}{4\pi^{2}}v_{1}-\mu^{2}\langle \hat{\phi}^{2}\rangle_{\subQ},
\end{align}
which yields the well-known trace anomaly for massless fields.

 \section{Numerical Implementation}
The main numerical task in the implementation of the prescriptions outlined in Sections \ref{sec:HH} and \ref{sec:B&U} is to obtain the radial modes by numerically integrating the radial equation for the  quantum state under consideration. The methods for treating the radial differential equation are quite different for the Euclidean modes and the Lorentzian modes, so we discuss each case separately.

\subsection{Euclidean Modes}
The two-point function on the Euclidean slice is obtained by the usual separation of variables procedure, the problem essentially reducing to a one-dimensional radial Green function $g_{nl}(r,r')$, where the mode numbers $n$ and $l$ are discrete. Solving this radial Green function amounts to computing a normalized product of solutions to the homogeneous equation (\ref{eq:Euclideanradialeqn}). The Hartle-Hawking state can be uniquely defined on this Euclidean slice by the condition of regularity on the event horizon. We denote the radial solution regular on the event horizon to be $p_{nl}(r)$ while the solution $q_{nl}(r)$ is the solution regular at infinity but divergent at the event horizon. To impose regularity of the Green function at the horizon, we must ensure that $p_{nl}(r)$ is evaluated at the smaller of the two radial points, i.e., we take $g_{nl}(r, r')=\kappa_+\,p_{nl}(r_{<})q_{nl}(r_{>})/N_{nl}$, where $r_{<}\equiv \textrm{min}\{r,r'\}$, $r_{>}=\textrm{max}\{r,r'\}$, and the factor of $\kappa_+$ is incorporated for convenience into our definition of $g_{nl}(r,r')$ so that it satisfies an equation with a $\kappa_+\,\delta (r-r')$ source term. The normalization constant comes from the  Wronskian $N_{nl}=-r^{2}f(r)\mathcal{W}\{p_{nl},q_{nl}\}$

All that remains is to compute the solutions $p_{nl}(r)$ and $q_{nl}(r)$. The computation of the former is simplified by recasting the radial equation into a confluent Heun form \cite{taylorbreenottewill:2021}. Since the confuent Heun functions that are regular at a regular singular point are built into many software suites such as Mathematica or Maple, computing these presents no difficulty. In particular, if we let $\mathsf{H}(q,\alpha, \gamma,\delta,\epsilon;z)$ be the confluent Heun function that solves
\begin{align}
    z(z-1) \mathsf{H}''(z)+ (\gamma (z-1)+\delta z+z(z-1)\epsilon) \mathsf{H}'(z)\nonumber\\
    +(\alpha z-q)\mathsf{H}(z)=0
\end{align}
that is analytic in the vicinity of $z=0$ and normalized to unity at $z=0$, then it is straightforward to show that:
\begin{align}
	p_{nl}(r)&=e^{-(\omega-\tilde{\omega})r}e^{\omega\,r_{*}}\mathsf{H}(q,\alpha,\gamma,\delta,\epsilon;z)
\end{align}
where 
\begin{align}
\label{eq:Heunparameters}
 q&=l(l+1)+r_{+}^{2}(\omega-\tilde{\omega})^{2}-(r_{+}+r_{-})(\omega-\tilde{\omega})-2\,r_{+}\tilde{\omega}\nonumber\\
 \alpha&=(r_{+}^{2}-r_{-}^{2})(\omega-\tilde{\omega})^{2}-2(r_{+}-r_{-})\tilde{\omega}\nonumber\\
 \gamma&=1+\frac{2\omega\,r_{+}^{2}}{r_{+}-r_{-}}\nonumber\\
 \delta&=1-\frac{2\omega\,r_{-}^{2}}{r_{+}-r_{-}}\nonumber\\
 \epsilon&=-2\tilde{\omega}(r_{+}-r_{-})\nonumber\\
 z&=\frac{r_{+}-r}{r_{+}-r_{-}},
 \end{align}
 and where in the expressions above we have used the notation
 \begin{align}
 \omega=n\,\kappa_+,\qquad \tilde{\omega}=\sqrt{n^{2}\kappa_+^{2}+\mu^{2}}.
\end{align}

Computing the $q_{nl}(r)$ modes is computationally harder. While these modes can still be written in confluent Heun form, the Heun functions with the appropriate boundary conditions for $q_{nl}(r)$ are not built into Mathematica or Maple. There are several options one can consider for computing $q_{nl}(r)$ but we found it most efficient to simply numerically integrate the radial equation inwards from a large $r$ value using an asymptotic expansion for the initial conditions. The initial conditions were optimized so that the asymptotic expansion solved the wave equation to our working precision with the least number of terms in the asymptotic expansion and for the smallest reasonable $r$ value at which this precision could be achieved. Using this approach, the mode solutions and their derivatives that were generated were accurate to at least 30 significant digits. We tested this accuracy by checking the constancy of the Wronskian over the radial grid for the solution pairs $\{p_{nl}(r),q_{nl}(r)\}$.

We wanted to compute the RSET to a high degree of accuracy which ostensibly requires a large set of $\{n,l\}$ modes. However, when employing the extended coordinates prescription as outlined in Section \ref{sec:HH}, the number of modes required can be significantly reduced by taking a suitably high order expansion of
the singular field. Here we choose to take a 6th order expansion (setting $m = 6$) in Eq. (\ref{eq:RegParam}) and generate 40 $l$ modes and 15 $n$ modes, which yields the RSET accurate to approximately 10-15 decimal places for the parameter sets considered in this paper.

\subsection{Lorentzian Modes}
In this section, we briefly describe the computation of the normalized Lorentzian modes $\{\Phi^{\textrm{in}}_{\omega l}(r),\Phi^{\textrm{up}}_{\omega l}(r)\}$. The boundary conditions on these modes are expressed in terms of transmission and reflection coefficients via (\ref{eq:LorentzianBCIn})-(\ref{eq:LorentzianBCUp}) with (\ref{eq:transmission1})-(\ref{eq:transmission2}). All that remains to compute is the relationship between the reflection/transmission coefficients and a pair of convenient numerically computed radial modes $\{\tilde{\Phi}_{\omega l}^{\textrm{in}}(r), \tilde{\Phi}_{\omega l}^{\textrm{up}}(r)\}$.

Starting first with the ``in'' modes. As in the Euclidean discussion above, the homogeneous solutions of the radial equation can be expressed in terms of confluent Heun functions, we take as $\tilde{\Phi}_{\omega l}^{\textrm{in}}(r)$ the solution
\begin{align}
    \tilde{\Phi}_{\omega l}^{\textrm{in}}(r)=\frac{1}{r_{+}}e^{-i\omega r_{*}}e^{i(\omega-\tilde{\omega})r}\mathsf{H}\left(q,\alpha,\gamma,\delta,\epsilon;z\right)
\end{align}
where here
\begin{align}
   \tilde{\omega}=\sqrt{\omega^{2}-\mu^{2}} 
\end{align}
and the parameters in the confluent Heun function are obtained from (\ref{eq:Heunparameters}) by applying the transformations \mbox{$\omega\to-i\omega$}, $\tilde{\omega}\to-i\tilde{\omega}$.
Comparing with the asymptotic forms, we see that
\begin{align}
    \tilde{\Phi}_{\omega l}^{\textrm{in}}(r)=\frac{e^{i(\omega-\tilde{\omega})r_{+}}}{B_{\omega l}^{\textrm{in}}}\Phi_{\omega l}^{\textrm{in}}(r).
\end{align}

As in the Euclidean case, we solve for $\tilde{\Phi}_{\omega l}^{\textrm{up}}(r)$ numerically by integrating the radial equation inwards from a suitably large radius with our initial conditions determined by an asymptotic series. The leading order term in this asymptotic series is
\begin{align}
    \tilde{\Phi}_{\omega l}^{\textrm{up}}(r)\sim \frac{e^{i\tilde{\omega} r_{*}}}{r},\qquad r\to\infty,
\end{align}
and the numerical solutions are related to the normalized modes by
\begin{align}
    \tilde{\Phi}_{\omega l}^{\textrm{up}}(r)=\frac{\Phi_{\omega l}^{\textrm{up}}(r)}{B_{\omega l}^{\textrm{up}}}.
\end{align}
With these definitions, the Wronskian condition implies
 \begin{align}
 B_{\omega l}^{\textrm{in}}=-\frac{2\,i\,\tilde{\omega}e^{i(\omega-\tilde{\omega})r_{+}}}{\mathcal{W}_{\omega l}},
 \end{align}
 where $\mathcal{W}_{\omega l}$ is the constant associated with the Wronskian of the numerical radial modes:
 \begin{align}
 \mathcal{W}_{\omega l}=r^{2}f\left(\tilde{\Phi}^{\textrm{up}}(r)\frac{d}{dr}\tilde{\Phi}^{\textrm{in}}(r) -\tilde{\Phi}^{\textrm{in}}(r)  \frac{d}{dr}\tilde{\Phi}^{\textrm{up}}(r)  \right).
 \end{align}

 The frequency integrals in the the RSET components have a rapid convergence so we compute modes on a frequency grid out to $\omega=10$. The error induced by truncating at this upper bound is very small. We choose a grid that is finely meshed near the lower limit of integration. For the $l$ sums appearing in (\ref{eq:RSETBoulware}), the convergence with $l$ is very rapid also, except for sums over the ``up'' modes very close to the horizon. Again because we desired to have very accurate results, we computed 150 $l$ modes for each frequency. The error induced by truncating the sums at $l=150$ is tiny for all but the points closest to the horizon. For example, for a massive field in the Reissner Nordstr\"{o}m spacetime with $Q/M=0.2$ and $\mu\, M=0.1$ (with $Q$ and $M$ denoting the electromagnetic charge and the ADM mass, respectively), for a fixed frequency, the error induced by truncating the $l$ sums at $l=150$ is less than 38 significant figures for points $(r-r_{+})/M\gtrapprox 0.05$.

 The last numerical issue we wish to briefly mention is that it is necessary to compute Lorentzian bound state modes with $\omega<\mu$. These are more difficult to compute than the modes with $\omega>\mu$. The main problem is that the asymptotic expansion does not converge to a sufficiently accurate value unless $r$ is chosen to be quite large whence the value of the initial conditions is usually very small. This forces one to increase the working precision of the numerical integrator and hence slows down the computation. The problem is accentuated for larger field masses. Fortunately, large field masses are not the physically relevant cases.

 \section{Results}\label{sec:results}
We have used the method previously outlined to obtain the RSET of massless and massive scalar fields in the Reissner-Nordström spacetime
\begin{align}\label{eq:ReissnerNordström}
    ds^{2}=
    &
    -\frac{\left(r-r_{+}\right)\left(r-r_{-}\right)}{r^{2}}dt^{2}+\frac{r^{2}}{\left(r-r_{+}\right)\left(r-r_{-}\right)}dr^{2}\nonumber\\
    &
    +r^{2}\left(d\theta^{2}+\sin{\theta}^{2}d\phi^{2}\right),
\end{align}
where $r_{\pm}=M\pm\sqrt{M^{2}-Q^{2}}$, in the
Hartle-Hawking, Boulware and Unruh vacuum states for different couplings and charge values. 
	
The RSET is an ambiguous quantity due to its dependence on the arbitrary lengthscale $\ell$ present in the 4-dimensional Hadamard parametrix.  
Similar to the Hadamard parametrix, such renormalization ambiguities are local and independent of the vacuum state under consideration. 
Consequently, all dependence on $\ell$ is already contained in $\langle\hat{T}^{a}{}_{b}\rangle_{\subHH}$. In spherical symmetry, Anderson, Hiscock and Samuel~\cite{AHS1995} (AHS) showed that $\ell$-dependent terms amount to a covariantly conserved analytic stress-tensor. Particularizing for the Reissner-Nordström spacetime~\eqref{eq:ReissnerNordström}, this contribution takes the form
	\begin{align}\label{eq:arbitraryren}
		\langle\hat{T}^{a}{}_{b}\rangle_{\ell}= \left(H^{a}{}_{b}+\mathcal{M}^{a}{}_{b}\right)\frac{\log\left(M/\ell\right)}{160\pi^{2}r^{8}},
	\end{align}
	with
	\begin{align}\label{eq:arbitraryterms}
		{H}^{t}{}_{t}=
		&
		3H^{r}{}_{r}=
		-\frac{3}{2}H^{\theta}{}_{\theta}=4\left(r-r_{+}\right)\left(r-r_{-}\right)r_{+}r_{-},\nonumber\\	
		\mathcal{M}^{r}{}_{r}=&\mathcal{M}^{t}{}_{t}=	5\mu^{4}r^{8}+20\mu^{2}\left(\xi-\frac{1}{6}\right)r^{4}r_{+}r_{-},\nonumber\\
		& \mathcal{M}^{\theta}{}_{\theta}=
		5\mu^{4}r^{8}-20\mu^{2}\left(\xi-\frac{1}{6}\right)r^{4}r_{+}r_{-},
	\end{align}
where clearly, for $\mu=0$, the terms~\eqref{eq:arbitraryterms} vanish at the event horizons for $Q\neq0$ and everywhere for $Q=0$.
 
The inherent ambiguities in the definition of the RSET can make comparisons between the results of different approaches to its calculation difficult. However, we are able to write down a relationship between results obtained via the AHS approach, which is based on Christensen's  DeWitt-Schwinger expansion \cite{Christensen:1976}, and the extended coordinate method, which is based on the Hadarmard parametrix. To do this we note that in addition to the ambiguity in the choice of lengthscale mentioned above, Mc Laughlin proved that, for a scalar field, an RSET obtained via Christensen's or by the Hadamard approach will differ by a geometric term, given by \cite{McLaughlin1990}:
  \begin{align}\label{eq:McL}
  	\langle\hat{T}_{ab}\rangle_{\textrm{McL}}=\frac{\mu^2}{16 \pi^2}\left[(\xi-\tfrac{1}{6})(R_{ab}-\tfrac{1}{2}Rg_{ab})-\tfrac{3}{8}\mu^2g_{ab}\right].
  \end{align}
  This is clearly a conserved quantity that can be absorbed into the semiclassical field equations via a renormalization of the constants $G$ and $\Lambda$.
    Therefore we have that the RSETs obtained via the two approaches are related by 
 	\begin{align}
 		\label{eq:compar}
 	\langle\hat{T}^{a}{}_{b}\rangle_{\textrm{AHS}}=		
  &
  \langle\hat{T}^{a}{}_{b}\rangle_{\textrm{EC}}+	\langle\hat{T}^{a}{}_{b}\rangle_{\textrm{McL}}\nonumber\\
  &
  + \left(H^{a}{}_{b}+\mathcal{M}^{a}{}_{b}\right)\frac{\log\left(\frac{e^{\gamma_{\subE}} \hat{\ell}}{2}\right)}{160\pi^{2}r^{8}}
 \end{align}
 where $\gamma_{\subE}$ is Euler's constant and  $\hat{\ell}=\ell_{\textrm{EC}}/\ell_{\textrm{AHS}}$ is the ratio of the choice of lengthscale made in both approaches. As the AHS approach is based on the DeWitt-Schwinger expansion, $\ell_{\textrm{AHS}}$ is conventionally taken to be equal to $1/\mu$ for a massive field and is arbitrary otherwise.  
 
 Eq. (\ref{eq:compar}) then enables meaningful comparisons between the two approaches, in particular in section~\ref{sec:comparison}, it will allow us to compare our exact results with the AHS and De-Witt Schwinger approximations to the RSET.

As the space of parameters to explore is large, in the next subsections we consider first how the different vacuum states affect the RSET in the massless and massive cases, to later analyse the impact of varying the coupling $\xi$ and the charge $Q$. We will compare these exact results with the values predicted by analytic RSET approximations in section~\ref{sec:comparison}.
	
	\subsection{Vacuum states}
\begin{figure*}
    \centering
    \includegraphics[width=\textwidth]{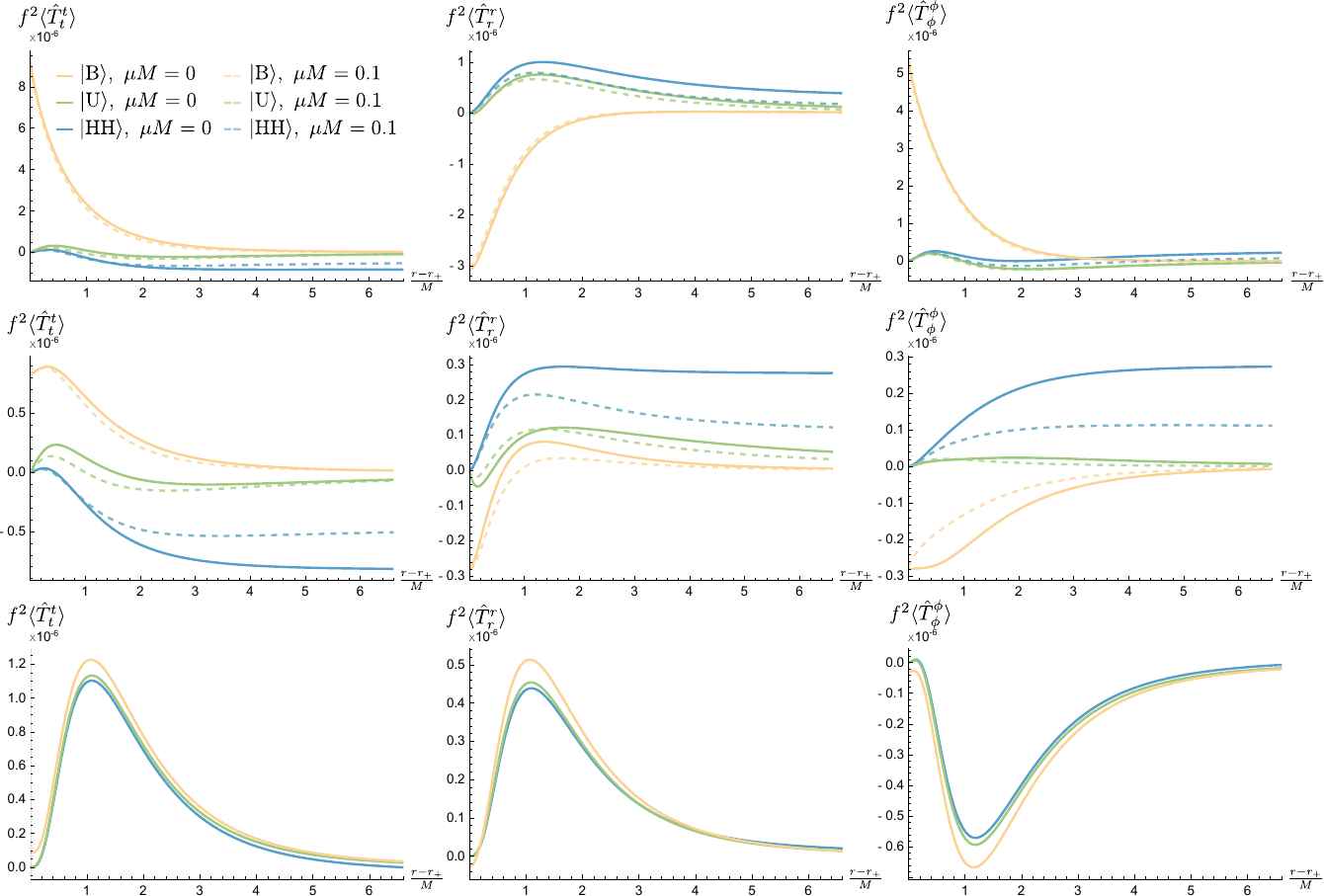}
    \caption{RSET components for various states, charge values and couplings. Continuous lines represent massless fields, while dashed lines denote fields with $\mu M = 0.1$. Blue, green and yellow curves denote RSET components in the Hartle-Hawking, Unruh and Boulware states, respectively. Left, middle and right columns each represent $f^{2}\langle\hat{T}^{t}{}_{t}\rangle$, $f^{2}\langle\hat{T}^{r}{}_{r}\rangle$ and $f^{2}\langle\hat{T}^{\phi}{}_{\phi}\rangle$. Top, middle and bottom rows each correspond to $\left\{\xi=0,~Q/M=0.2\right\}$, $\left\{\xi=1/6,~Q/M=0.2\right\}$ and $\left\{\xi=1/6,~Q/M=0.99\right\}$.}
    \label{fig:states}
\end{figure*}

	Figure~\ref{fig:states} contains plots of the RSET components multiplied by $f^{2}$ for $\left\{\xi=0,~Q/M=0.2\right\}$ and \mbox{$\left\{\xi=1/6,~Q/M=0.2\right\}$}, both with $\mu M=0$ (continuous lines) and $\mu M = 0.1$ (dashed lines), in the first and second rows, respectively. The third row depicts the near-extremal case $\left\{\xi=1/6,~Q/M=0.99\right\}$ with $\mu M = 0$. For every coupling, charge, and mass, all RSET components in the Hartle-Hawking states are finite at the event horizon (strictly, they are finite at the lowest point in our radial grid, which can be taken as close to $r=r_{+}$ as desired, however the regularity of all components on the horizon was proven in \cite{BreenOttewill2012a}). 
	As expected, every RSET component in the Boulware state diverges at the horizon in a way \mbox{$\propto f^{-2}$} for all the couplings considered, while for the Unruh state  $\langle\hat{T}^{r}{}_{r}\rangle$ and $\langle\hat{T}^{t}{}_{t}\rangle$ diverge like  \mbox{$\propto f^{-1}$}.   For large $r$, the diagonal RSET components in the Unruh and Boulware states approach the same value, in accordance with the asymptotic behaviours found in~\cite{ChristensenFulling1977,Candelas1980}. 
	We also find that in a freely falling frame, as expected,  $\langle\hat{T}^{a}{}_{b}\rangle$ is regular on the future event horizon for the Unruh state, regular on the future and past event horizons for the Hartle-Hawking state, and diverges on both the past and future horizons in the Boulware state.

	Increasing the mass of the field affects the value and sign of the RSET at large $r$. In the Boulware and Unruh states, these do not decay to zero asymptotically, as with $\mu M = 0$, but approach a constant value instead. These asymptotic values can be identified with the choice of arbitrary length scale. In fact, from the numerical results we find that the RSET for the Boulware and Unruh states approach the following value
	 \begin{equation}
		\langle\hat{T}^{a}{}_{b}\rangle_{\infty}=\frac{\delta^{a}{}_{b}\mu^{4}}{128\pi^{2}}\left[3+4\log\left(\frac{2e^{-\gamma_{\subE}}}{\mu\, \ell_{\textrm{EC}}}\right)\right].
	\end{equation}
	From inspection of Eq (\ref{eq:compar}), we see that this corresponds to the large $r$ expansion of the difference between the extended coordinates and AHS RSETs (with $\ell_{\textrm{AHS}}=1/\mu$). For massive fields, we therefore have a natural choice for the lengthscale $\ell_{\text{EC}}$, for which the Boulware and Unruh RSET decays to 0 as $r \to \infty$, given by 
	\begin{align}
	\ell_{\text{EC}}	=\frac{2}{\mu} \mbox{exp}\{3/4-\gamma_{\subE}\}.
		\end{align}
	We make this choice for the massive field results presented in this Section. For the massless field, the large $r$ asymptotic behaviour is independent of the choice of lengthscale, so for simplicity, in this case we set \mbox{$\ell_{EC}=M$}, where $M$ is the black hole mass. 

The aforementioned properties hold for all $Q<M$. As $Q$ increases, the finite terms relating the various states in Eq.~\eqref{eq:RSETBoulware} decrease in magnitude due to their dependence on the black hole temperature, and the Boulware, Unruh and Hartle-Hawking states converge. The Unruh and Hartle-Hawking states converge faster than the Boulware state, since contributions from the ``up" modes to Eq.~\eqref{eq:RSETBoulware} decrease slower in the $Q\to M$ limit than contributions from the ``in" modes. Results for the extremal case $Q=M$ will appear elsewhere.

\subsection{Varying the coupling}
\begin{figure*}
        \centering
        \includegraphics[width=\textwidth]{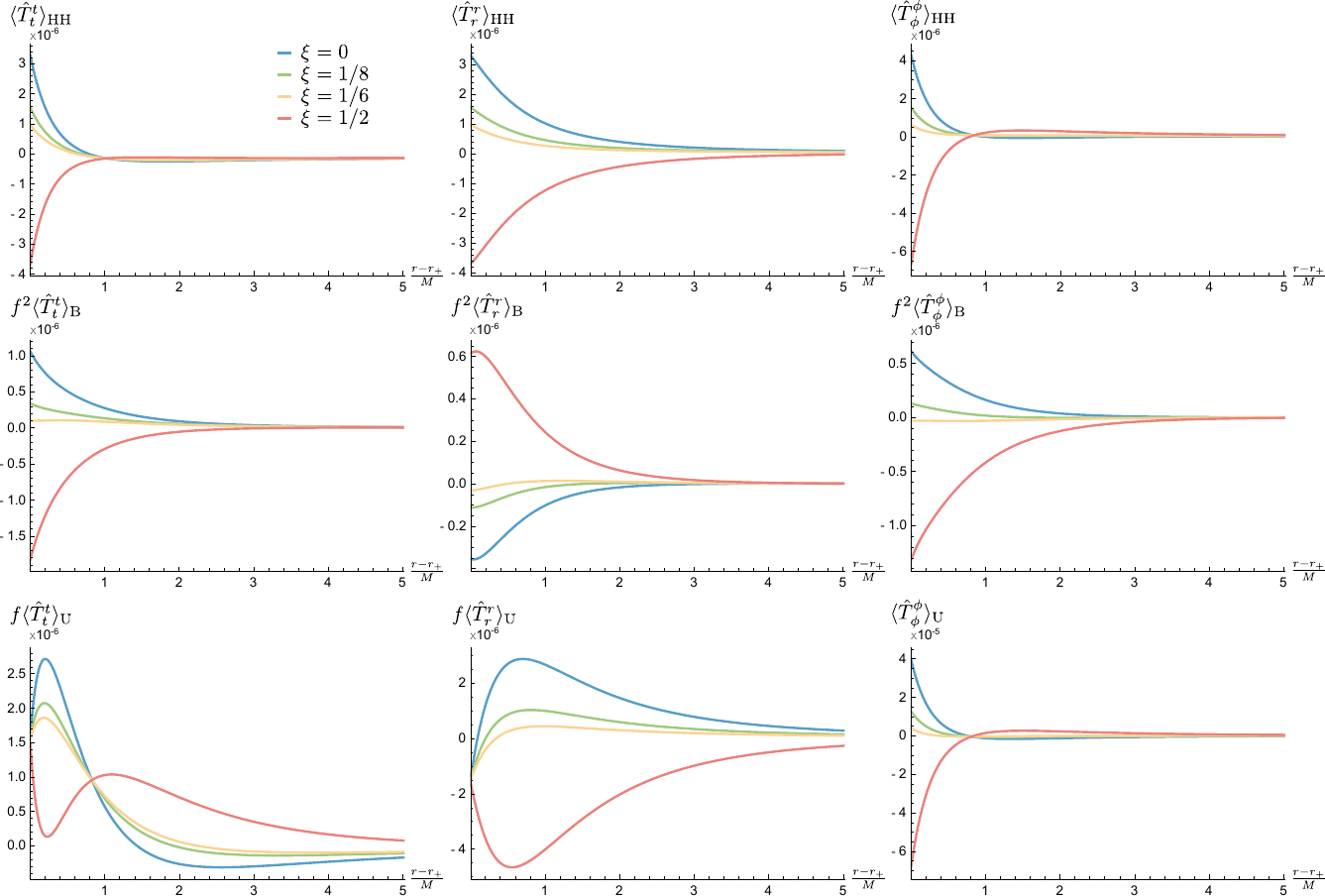}
        \caption{RSET in terms of $r/M$ for various couplings and vacuum states for the Reissner-Nordström black hole with $Q/M=0.6$. The blue, green, yellow and orange curves correspond to the couplings $\xi=\left\{0,1/8,1/6,1/2\right\}$, respectively. Left, middle and right columns each represent 
         $\langle\hat{T}^{t}{}_{t}\rangle$, $\langle\hat{T}^{r}{}_{r}\rangle$ and $\langle\hat{T}^{\phi}{}_{\phi}\rangle$. Top, middle and bottom rows correspond to the Hartle-Hawking, Boulware and Unruh states.}
        \label{fig:couplings}
\end{figure*}

Next we consider the effect of varying the field coupling $\xi$. Figure~\ref{fig:couplings} shows the diagonal RSET components in the Hartle-Hawking, Boulware and Unruh states (first, second and third row, respectively) for $Q/M=0.6$. Components in the Boulware and Unruh states have been multiplied by powers of $f$ for visualization purposes. 

Increasing the value of the coupling does not affect notably the magnitude of the RSET, but it does produce a change in the sign of every RSET component with the exception of $\langle\hat{T}^{t}{}_{t}\rangle_{\subU}$ and $\langle\hat{T}^{r}{}_{r}\rangle_{\subU}$. 
As was shown in~\cite{taylorbreenottewill:2021} for the Hartle-Hawking state, RSET values for any coupling can be generated from results in minimal coupling (this follows from definition~\eqref{eq:RSET} in spacetimes with vanishing Ricci scalar). With this we find that, for $Q/M=0.6$, the $\langle\hat{T}^{t}{}_{t}\rangle_{\subHH}$ and $\langle\hat{T}^{r}{}_{r}\rangle_{\subHH}$ components change sign at the lowest grid point at \mbox{$\xi\approx 0.24$} and the $\langle\hat{T}^{\theta}{}_{\theta}\rangle_{\subHH}$ component component does at \mbox{$\xi\approx0.19$}. For the first two components, said sign changes occurs at smaller $\xi$ as $Q$ increases, whereas the converse happens to the latter component. For the Boulware state, the analytic RSET approximation from ~\cite{AHS1995} suggests that the sign of the RSET at the horizon is independent from $Q$ (see Sec.~\ref{sec:comparison} below).

For diagonal RSETs, the point-wise null energy condition is satisfied as long as the inequalities 
\begin{equation}
    -\langle\hat{T}^{t}{}_{t}\rangle+\langle\hat{T}^{r}{}_{r}\rangle\geq0,\quad -\langle\hat{T}^{t}{}_{t}\rangle+\langle\hat{T}^{\phi}{}_{\phi}\rangle\geq0,
\end{equation}
hold everywhere outside the event horizon~\cite{Visser1996a,Visser1996b}. The Hartle-Hawking state satisfies the null energy condition everywhere outside the event horizon for \mbox{$\xi=\left\{0,1/8,1/6\right\}$}, while for $\xi=1/2$ it is violated from $r\approx 2.8 r_{+}$ to $r=r_{+}$. The Boulware state violates the null energy condition everywhere outside and at the event horizon for 
\mbox{$\xi=\left\{0,1/8,1/6\right\}$} and satisfies it everywhere for $\xi=1/2$. Note that these behaviours depend on the particular choice of renormalization lengthscale~\eqref{eq:arbitraryren} and hence should not be considered to be physically meaningful statements. 

\subsection{Varying the charge}
\begin{figure*}
        \centering
        \includegraphics[width=\textwidth]{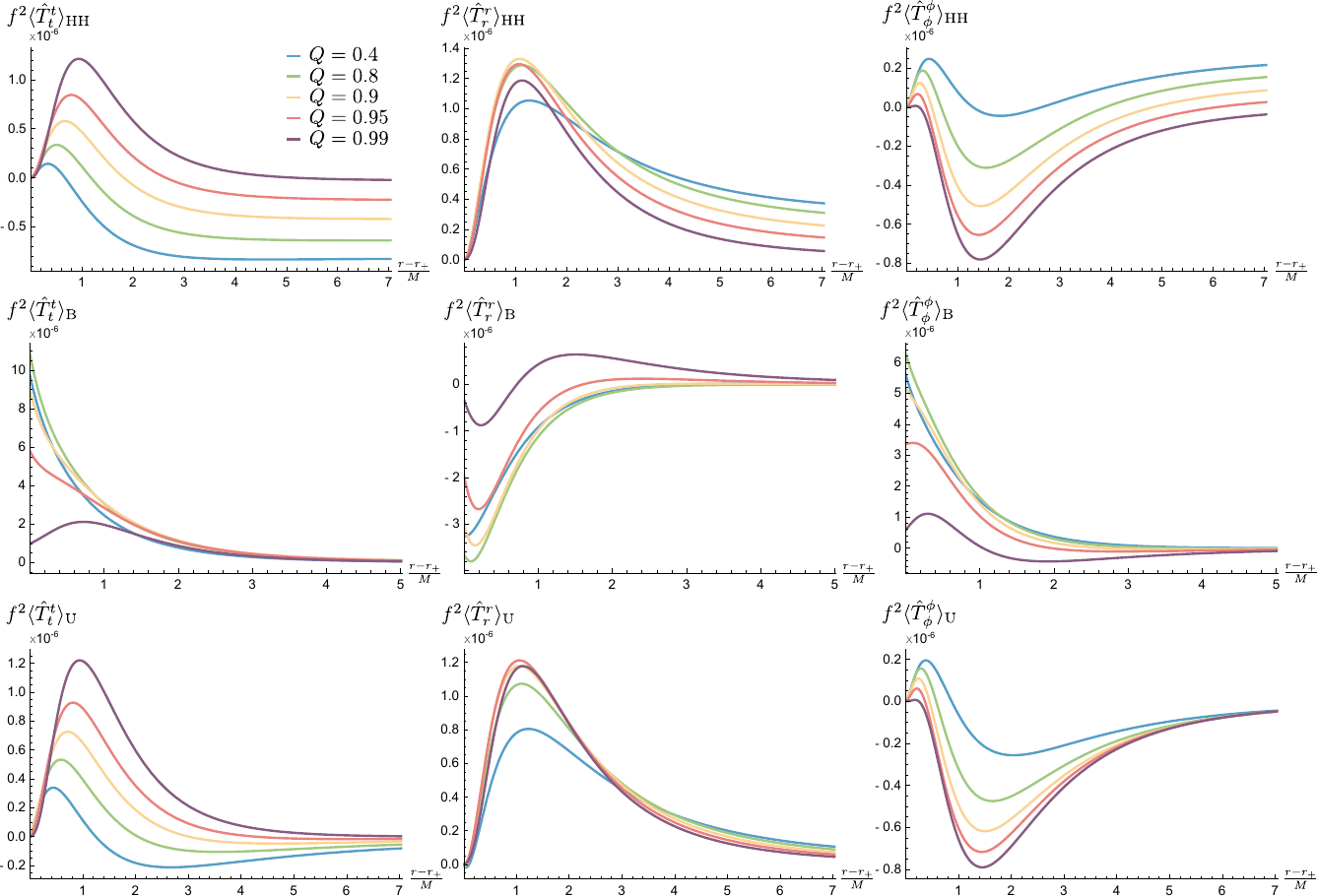}
        \caption{RSET components in the Hartle-Hawking, Boulware, and Unruh states (first, second, and third rows, respectively) with $\xi=0$. The blue, green, yellow, red and purple curves correspond to $Q/M=\left\{0.4,0.8,0.9,0.95,0.99\right\}$, respectively. Left panel is 
        $\langle\hat{T}^{t}{}_{t}\rangle$, middle panel is  $\langle\hat{T}^{r}{}_{r}\rangle$ and right panel is $\langle\hat{T}^{\phi}{}_{\phi}\rangle$.}
        \label{fig:Charges}
\end{figure*}

Varying the black hole charge $Q$ affects both the sign and the magnitude of the RSET components, which become more sensitive to slight variations of $Q$ as the extremal limit $Q=M$ is approached. Figure~\ref{fig:Charges} shows the RSET components for $\xi=0$ and the charge values \mbox{$Q/M=\left\{0.4,0.8,0.9,0.95,0.99\right\}$}. We study the minimally coupled case in detail to allow for a faithful comparison with the analytic approximations presented in Section~\ref{sec:comparison}.

For the RSET in the Hartle-Hawking state, increasing $Q$ decreases the asymptotic values in every RSET component at large $r$. Asymptotically, this state reproduces a thermal bath at equilibrium with the horizon temperature, which decreases with increasing $Q$. At the horizon $r=r_{+}$, the components $\langle\hat{T}^{t}{}_{t}\rangle_{\subHH}$ and $\langle\hat{T}^{r}{}_{r}\rangle_{\subHH}$ increase with $Q$ except for $Q/M=0.99$, for which these components are smaller than in the $Q/M=0.95$ case. For $\langle\hat{T}^{\phi}{}_{\phi}\rangle_{\subHH}$, this inversion happens at lower $Q$ and is already noticed in the $Q/M=0.95$ case, even becoming negative at the event horizon for $Q/M=0.99$. 

In the Boulware state, the sign of the RSET components does not change at both extremes of our radial grid or anywhere in between, except for a narrow  region in the $Q/M=0.99$ case. As the charge is increased, the positive coefficient that controls the divergence of the RSET at $r=r_{+}$ decreases, but this does not change the fact that every component diverges $\propto f^{-2}$, which only become finite at the horizon in the extremal case. 

For the Unruh state, again the sign of the RSET components is independent of $Q$, but their overall magnitude diminishes with the charge, in consistency with this state describing a flux of black-body radiation for large $r$. At the event horizon, the $\langle\hat{T}^{t}{}_{t}\rangle_{\subU}$ and $\langle\hat{T}^{r}{}_{r}\rangle_{\subU}$ components decrease and increase with $Q$, respectively, whereas the $\langle\hat{T}^{\phi}{}_{\phi}\rangle_{\subU}$ component increases for $Q/M=\left\{0.4,0.8,0.9\right\}$, decreases with $Q/M=0.95$, and becomes negative for $Q/M=0.99$, as in the Hartle-Hawking state. 

In the extremal case $Q=M$, it was shown in~\cite{Andersonetal1995} that the RSET of massless fields at the event horizon takes the same value as the RSET of a conformally invariant field in the Bertotti-Robinson spacetime~\cite{OttewillTaylor2012}. The magnitude of the RSET at $r=r_{+}$ varies more abruptly for $Q/M>0.9$, hence we do not observe they approach the values obtained in the Bertotti-Robinson spacetime for $Q/M=0.99$. To observe this tendency, $Q$ values even closer to $M$ would need to be explored.

\section{Comparison with Approximations}
\label{sec:comparison}
Obtaining accurate results for the RSET, even in scenarios of high symmetry like the Reissner-Nordström black hole, proves to be a computationally expensive task: a large amount of highly precise Euclidean and Lorentzian field modes need to be calculated for each collection of parameters $\left\{M,Q,\xi,\mu\right\}$. Once results are available for the RSET, we can find its backreaction on the background metric at $\mathcal{O}\left(\hbar\right)$ through Eqs.~\eqref{eq:semieqs}. If one insists in progressing along this line to find the backreacted metric at higher-orders in $\hbar$, it is necessary to compute, every time, new sets of modes propagating over the backreacted metric. Such a scheme proves to be unreasonably time consuming.

Analytic RSET approximations alleviate the difficulties behind computing the RSET and its backreaction since they bypass numerical mode calculations, expressing the RSET (almost) exclusively in terms of the metric components and their derivatives. This simplifies dramatically the complexity of semiclassical analyses at the cost of reducing the physical content encoded within the RSET itself. Furthermore, there is no unique or preferred approximate scheme available in the literature. Instead, we have multiple RSET approximations based on different physical principles. Despite this non-uniqueness,  
analytic RSETs should nonetheless reproduce the physics of their exact counterparts at least qualitatively, replicating the defining properties of the different vacuum states and yielding correct results at the asymptotic regions of the spacetime. In this section, we will review the various RSET approximations available in spherical symmetry and compare them with the exact results presented in Section~\ref{sec:results}.

\subsection{Analytic RSETs in spherical symmetry}


\subsubsection{The Polyakov RSET}
In spherical symmetry, we can obtain various analytic RSETs by fixing the field parameters of the theory, namely, the field mass and the coupling. Perhaps the most well-known example is that of conformally invariant fields $\left\{\mu=0,\xi=1/6\right\}$ in conformally flat backgrounds, where the RSET is fully determined by the local trace anomaly~\cite{BrownCassidy1977}. In more generic spherically symmetric spacetimes, the essential features of the propagation of a massless minimally coupled scalar $\left\{m=0,\xi=0\right\}$ in four spacetime dimensions can be captured by two-dimensional models, described by the line element
\begin{equation}\label{eq:2Dmetric}
ds^{2}_{\text{(2D)}}=-f(r)dt^{2}+dr^{2}/f(r).
\end{equation}
This connection between $4$D and $2$D physics manifests upon taking the near-horizon limit $r\to r_{+}$ in Eq.~\eqref{eq:Schrodinger} for the $s$-wave ($l=0$) component of the field. 
The gravitational potential vanishes in this limit, and the $(t,r)$ sector reduces to the two-dimensional free wave equation, which is conformally invariant (to see this explicitly, we avoid expanding in $t$ in~\eqref{eq:harmsph}, see~\cite{FabbriNavarro-Salas2005} for details). The emergence of this symmetry allows to express the $2$D RSET in closed analytic form~\cite{DaviesFulling1977}. These $2$D expressions are then identified with a $4$D RSET through
\begin{equation}\label{eq:2Dto4D}
\langle\hat{T}^{a}{}_{b}\rangle^{\rm P}=\frac{1}{4\pi r^{2}}\delta^{a}{}_{A}\delta^{B}{}_{b}\langle\hat{T}^{A}{}_{B}\rangle^{\text{(2D)}},
\end{equation}
where $\{a,b\}$ run over the 4D spacetime indices while $\{A,B\}$ run over the 2D spacetime indices, and $\rm P$ stands for the Polyakov RSET. The multiplicative factor $1/4\pi r^{2}$ has been introduced to ensure covariant conservation in $4$D and so that this approximation reproduces the adequate Unruh fluxes at infinity [see Eq.~\eqref{eq:HawkingFlux} below]. In black hole spacetimes, the components are
\begin{align}\label{eq:polyakovRSET}
\langle\hat{T}^{t}{}_{t}\rangle^{\rm P}=
&
\frac{1}{384\pi^{2}r^{2} f}\left[\left(f'\right)^{2}-4f f''\right]+\langle\hat{T}^{t}{}_{t}\rangle_{\kappa_+}^{\rm P},\nonumber\\
\langle\hat{T}^{r}{}_{r}\rangle^{\rm P}=
&
-\frac{1}{384\pi^{2}r^{2} f}\left(f'\right)^{2}+\langle\hat{T}^{r}{}_{r}\rangle_{\kappa_+}^{\rm P},
\end{align}
where \mbox{$f=(r-r_{+})(r-r_{-})/r^{2}$} in the Reissner-Nordström spacetime and angular pressures vanish. The terms $\langle\hat{T}^{a}{}_{b}\rangle_{\kappa_+}^{\rm P}$ are temperature-dependent and relate RSETs in different vacuum states in a manner analogous to the integrals in Eqs.~\eqref{eq:RSETBoulware}. Here, they amount to the Schwarzian derivative between null coordinates in two different conformal coordinate systems~\cite{Fabbrietal2003,FabbriNavarro-Salas2005}. The Boulware state is defined as the state for which such temperature-dependent terms vanish. For the Hartle-Hawking state they equal
\begin{align}\label{eq:Polyakovtemp}
\langle\hat{T}^{t}{}_{t}\rangle_{\kappa_+, \subHH}^{\rm P}=
&
-\langle\hat{T}^{r}{}_{r}\rangle_{\kappa_+, \subHH}^{\rm P}=-\frac{\kappa_+^{2}}{96\pi^{2}r^{2} f},
\end{align}
where $\kappa_+=\left(r_{+}-r_{-}\right)/\,2\,r_{+}^{2}$. For the Unruh state we find
\begin{align}\label{eq:UnruhPolyakov}
\langle\hat{T}^{a}{}_{b}\rangle_{\kappa_+, \subU}^{\rm P}=\frac{1}{2}\langle\hat{T}^{a}{}_{b}\rangle_{\kappa_+, \subHH}^{\rm P}
\end{align}
and 
\begin{equation}\label{eq:HawkingFlux}
\langle\hat{T}_{t}{}^{r}\rangle_{\kappa_+, \subU}^{\rm P}=-\frac{\kappa_+^{2}}{192\pi^{2}r^{2}}.
\end{equation}
At spatial infinity, this component describes the usual Hawking flux emitted by an evaporating black hole. It can be easily checked that~\eqref{eq:Unruhflux} reduces to the above expression in the massless case by ignoring backscattering ($|B_{\omega l}^{\text{in}}|=1$) and neglecting $l>0$ multipoles.

As it properly describes black hole evaporation, the Polyakov approximation has been extensively used in the literature~\cite{ParentaniPiran1994,AyalPiran1997,Chakrabortyetal2015,Arrecheaetal2021}. Note that the Polyakov RSET is ill-defined at $r=0$. This pathology motivates the search for regularized Polyakov RSETs that display a regular behaviour at $r=0$~\cite{Arrecheaetal2021} while reproducing~\eqref{eq:HawkingFlux} at large distances.

\subsubsection{The s-wave RSET}
It is possible to incorporate into the Polyakov RSET the backscattering effects of the gravitational potential (which we had neglected via the near-horizon approximation) by considering a two-dimensional scalar coupled to a dilaton field (we refer the reader to~\cite{FabbriNavarro-Salas2005} for details on this approach). This method is hybrid, in the sense that the resulting $2$D RSET is non-conserved, such non-conservation being identified as the angular components of the following $4$D RSET,
\begin{align}\label{eq:swaveRSET}
\langle\hat{T}^{t}{}_{t}\rangle^{\rm s}=
&
\langle\hat{T}^{t}{}_{t}\rangle^{\rm P}+\langle\hat{T}^{t}{}_{t}\rangle_{\kappa_+}^{\rm s}\nonumber\\
&
-\frac{\left(r-r_{+}\right)\left(r-r_{-}\right)}{32\pi^{2}r^{6}}\log\frac{\left(r-r_{+}\right)\left(r-r_{-}\right)}{r^{2}},\nonumber\\
\langle\hat{T}^{r}{}_{r}\rangle^{\rm s}=
&
\langle\hat{T}^{r}{}_{r}\rangle^{\rm P}+\langle\hat{T}^{r}{}_{r}\rangle_{\kappa_+}^{\rm s}\nonumber\\
&
+\frac{\left(r-r_{+}\right)\left(r-r_{-}\right)}{32\pi^{2}r^{6}}\log\frac{\left(r-r_{+}\right)\left(r-r_{-}\right)}{r^{2}},
\nonumber\\
&
+\frac{2r_{+}r_{-}-r\left(r_{+}+r_{-}\right)}{16\pi^{2}r^{6}},\nonumber\\
\langle\hat{T}^{\phi}{}_{\phi}\rangle^{\rm s}=
&
\langle\hat{T}^{\phi}{}_{\phi}\rangle_{\kappa_+}^{\rm s}-\left[22r_{+}^{2}r_{-}^{2}-29rr_{+}r_{-}\left(r_{+}+r_{-}\right)\right.\nonumber\\
&
\left.+2r^{2}\left(r_{+}+4r_{-}\right)\left(4r_{+}+4_{-}\right)-7r^{3}\left(r_{+}+r_{-}\right)\right]\nonumber\\
&
\times\left[64\pi^{2}r^{6}\left(r-r_{+}\right)\left(r-r_{-}\right)\right]^{-1}\nonumber\\
&
+\frac{\left[r^{2}-2r\left(r_{+}+r_{-}\right)+3r_{+}r_{-}\right]}{32\pi^{2}r^{6}}\times\nonumber\\
&
\log\frac{\left(r-r_{+}\right)\left(r-r_{-}\right)}{r^{2}},
\end{align}
where $4$D time and radial components have been obtained through relation~\eqref{eq:2Dto4D}. The temperature-dependent terms now acquire the more involved form
\begin{align}\label{eq:swavetemp}
\langle\hat{T}^{t}{}_{t}\rangle_{\kappa_+, \subHH}^{\rm s}=
&
-\langle\hat{T}^{r}{}_{r}\rangle_{\kappa_+, \subHH}^{\rm s}=\langle\hat{T}^{t}{}_{t}\rangle_{\kappa_+, \subHH}^{\rm P}+\nonumber\\
&
\frac{\kappa_+}{16\pi^{2} r^{6}}\left\{\vphantom{\log\left(\frac{r-r_{-}}{r_{-}}\right)}r^{2}\left(r_{+}+r_{-}\right)-r\,r_{+}r_{-}\right.\nonumber\\
&
\left.-\frac{\left(r-r_{+}\right)\left(r-r_{-}\right)}{r_{+}-r_{-}}\times\right.\nonumber\\
&
\left.\left[r_{+}^{2}\log\left(\frac{r-r_{+}}{r_{+}}\right)-r_{-}^{2}\log\left(\frac{r-r_{-}}{r_{-}}\right)\right]\right\},\nonumber\\
\langle\hat{T}^{\phi}{}_{\phi}\rangle_{\kappa_+, \subHH}^{\rm s}=
&
\frac{\kappa_+}{32\pi^{2}r^{6}}\times\left\{\frac{r}{\left(r-r_{+}\right)\left(r-r_{-}\right)}\right.\times\nonumber\\
&
\left.\left[3r^{3}\left(r_{+}+r_{-}\right)-4r^{2}\left(r_{+}^{2}+3r_{+}r_{-}+r_{-}^{2}\right)\right.\right.\nonumber\\
&
\left.\left.+10r_{+}r_{-}r\left(r_{+}+r_{-}\right)-6r_{+}^{2}r_{-}^{2}\right]\right.\nonumber\\
&
\left.+\left[4r-\frac{2r^{2}+6r_{+}r_{-}}{r_{+}-r_{-}}\right]\times\right.\nonumber\\
&
\left.\left[r_{+}^{2}\log\left(\frac{r-r_{+}}{r_{+}}\right)-r_{-}^{2}\log\left(\frac{r-r_{-}}{r_{-}}\right)\right]\right\}
\end{align}
for the Hartle-Hawking state. For the Unruh state, these terms acquire a dependence on the time coordinate $t$ and the resulting RSET is not covariantly conserved.  

\subsubsection{The Anderson-Hiscock-Samuel RSET}
Dimensional reduction is not the only procedure to derive analytic RSET approximations. In $4$D, there is the analytic RSET approximation derived by Anderson, Hiscock and Samuel~\cite{AHS1995} that incorporates the effects of field mass and curvature coupling. This approximation naturally arises from point-splitting regularization, which results in a separation of the exact RSET into two independently conserved analytic and numeric parts. The analytic portion ---or AHS-RSET hereafter--- gives the correct trace anomaly in the $\left\{m=0,\xi=1/6\right\}$ case. This approximation yields a well-behaved RSET at $r=0$, but in exchange it exhibits third and fourth order derivatives of the metric functions and depends on the arbitrary lengthscale $\ell$. Expressions for this RSET, which we avoid showing here since they are lengthy and opaque, can be found in~\cite{AHS1995}. The AHS-RSET can describe the Boulware and Hartle-Hawking states, but it cannot describe the Unruh state.

As was hinted in~\cite{AHS1995}, the AHS-RSET is not an appropriate approximation for massive fields: it does not reproduce standard results in flat spacetime. Upon evaluating the AHS-RSET in Minkowski spacetime we obtain
\begin{align}\label{eq:AHSMink}
    \langle\hat{T}^{r}{}_{r}\rangle^{\rm AHS}_{\subHH}=
    &
    \frac{\kappa_+^{4}}{1440\pi^{2}}-\frac{\kappa_+^{2}\mu^{2}}{96\pi^{2}}
    -\frac{\mu^{4}}{128\pi^{2}}\left[4\log\left(\lambda\ell\right)+3\right],\nonumber\\
    \langle\hat{T}^{t}{}_{t}\rangle^{\rm AHS}_{\subHH}=
    &
    -\frac{\kappa_+^{4}}{480\pi^{2}}+\frac{\kappa_+^{2}\mu^{2}}{96\pi^{2}}
    -\frac{\mu^{4}}{128\pi^{2}}\left[4\log\left(\lambda\ell\right)-1\right],
\end{align}
with $\langle\hat{T}^{\theta}{}_{\theta}\rangle^{\rm AHS}=\langle\hat{T}^{\phi}{}_{\phi}\rangle^{\rm AHS}=\langle\hat{T}^{r}{}_{r}\rangle^{\rm AHS}$, and $\lambda$ is a positive parameter related to an infrared cutoff in some integrals from~\cite{AHS1995}. In the Boulware state, $\lambda$ can be absorbed in $\ell$ if the field is massless (being an arbitrary parameter otherwise), whereas in the Hartle-Hawking state, $\lambda=\kappa_+ \,\text{exp}\left(-\gamma_{E}\right)$. In view of~\eqref{eq:AHSMink}, we cannot make the AHS-RSET amount to a $4$D thermal bath with temperature $\kappa_{+}/2\pi$ in flat spacetime by an appropriate choice of $\ell$. Neither can we identify these components with a renormalization of the cosmological constant, as we did in section~\ref{sec:results} for the exact RSET in the Boulware state. Therefore, we regard the AHS-RSET as an inadequate approximation for massive fields, and consider $\mu=0$ hereafter when referring to this approximation.

As the reader may have noticed from~\eqref{eq:AHSMink}, the AHS-RSET contains an explicit dependence on the temperature. In the Reissner-Nordström spacetime, these terms take the form
\begin{align}\label{eq:AHStemp}
    \langle\hat{T}^{t}{}_{t}\rangle_{\kappa_+}^{\rm AHS}=
    &
    -\frac{\kappa_+^{4}}{480\pi^{2}f^{2}}+\frac{\kappa_+^{2}\left(\xi-\frac{1}{6}\right)}{32\pi^{2}r^{4}f^{2}}\times\nonumber\\
    &
    \left[2r_{+}^{2}r_{-}^{2}-2r\,r_{+}r_{-}\left(r_{+}+r_{-}\right)+r^{2}\left(r_{+}^{2}+r_{-}^{2}\right)\right],\nonumber\\
    \langle\hat{T}^{r}{}_{r}\rangle_{\kappa_+}^{\rm AHS}=
    &
    \frac{\kappa_+^{4}}{1440\pi^{2}f^{2}}+\frac{\kappa_+^{2}\left(\xi-\frac{1}{6}\right)}{96\pi^{2}r^{4}f^{2}}\times\nonumber\\
    &
    \left[2r_{+}^{2}r_{-}^{2}-6r\,r_{+}r_{-}\left(r_{+}+r_{-}\right)\right.\nonumber\\
    &
    \left.+3r^{2}\left(r_{+}^{2}+4r_{+}r_{-}+r_{-}^{2}\right)-4r^{3}\left(r_{+}+r_{-}\right)\right],\nonumber\\
    \langle\hat{T}^{\phi}{}_{\phi}\rangle_{\kappa_+}^{\rm AHS}=
    &
    \frac{\kappa_+^{4}}{1440\pi^{2}f^{2}}+\frac{\kappa_+^{2}\left(\xi-\frac{1}{6}\right)}{48\pi^{2}r^{4}f^{2}}\times\nonumber\\
    &
    \left[r_{+}^{2}r_{-}^{2}-3r^{2}r_{+}r_{-}+r^{3}\left(r_{+}+r_{-}\right)\right].
\end{align}
Notice how these terms differ from those given by the Polyakov~\eqref{eq:Polyakovtemp} and $s$-wave approximations~\eqref{eq:swavetemp}. These differences will have a major impact on the behaviour of approximate RSETs at the event horizon on the different vacuum states.

\subsubsection{The DeWitt-Schwinger RSET}
The work~\cite{AHS1995} also derived expressions for the RSET in the DeWitt-Schwinger approximation for fields of large mass. The corresponding expressions [obtained to $\mathcal{O}\left({m^{-2}}\right)$] are local, have no information about the state of the vacuum, and can be used as an approximation to the complete RSET in situations where the field mass is comparable to the ADM mass $M$.

The components of the DS-RSET are
\begin{align}\label{eq:DS-RSET}
40320\pi^{2}\mu^{2}r^{12}\langle\hat{T}^{t}{}_{t}\rangle^{\rm DS}=
&
\sum^{4}_{i=0} A_{i}r^{i}+\xi\sum^{4}_{i=0} B_{i}r^{i},\nonumber\\
40320\pi^{2}\mu^{2}r^{12}\langle\hat{T}^{r}{}_{r}\rangle^{\rm DS}=
&
\sum^{4}_{i=0} C_{i}r^{i}+\xi\sum^{4}_{i=0} D_{i}r^{i},\nonumber\\
40320\pi^{2}\mu^{2}r^{12}\langle\hat{T}^{\phi}{}_{\phi}\rangle^{\rm DS}=
&
\sum^{4}_{i=0} E_{i}r^{i}+\xi\sum^{4}_{i=0}F_{i}r^{i},
\end{align}
where the coefficients in the sums are given in Eq.~\eqref{eq:DS-Coefs} in the Appendix.

In the next subsection we explore the accuracy of the aforementioned RSET approximations in the Reissner-Nordström black hole spacetime.

\subsection{Comparing exact and approximate RSETs}
Establishing comparisons between 
RSET approximations is complicated by the presence of renormalization ambiguities. For the sake of brevity, we will focus on the features of the RSET that are independent of these ambiguities. These are: the behaviour at the event horizon in the different states where, for massless fields, the RSET is independent of $\ell$ [see Eq.~\eqref{eq:arbitraryren}]; the form of temperature-dependent terms, and the trace anomaly.

\subsubsection{The Hartle-Hawking state and Unruh states}
In the Hartle-Hawking state we find the following leading-order contributions for the Polyakov RSET at the event horizon,
\begin{align}
\langle\hat{T}^{t}{}_{t}\rangle^{\rm P}_{\subHH}=
\langle\hat{T}^{r}{}_{r}\rangle^{\rm P}_{\subHH}=
\frac{r_{+}-2r_{-}}{96\pi^{2}r_{+}^{5}}+\mathcal{O}\left(r-r_{+}\right),
\end{align}
with vanishing angular pressures,
whereas for the $s$-wave RSET
\begin{align}
\langle\hat{T}^{t}{}_{t}\rangle^{\rm s}_{\subHH}=
&
\langle\hat{T}^{r}{}_{r}\rangle^{\rm s}_{\subHH}=
-\frac{2r_{+}-r_{-}}{96\pi^{2}r_{+}^{5}}+\mathcal{O}\left(r-r_{+}\right),\nonumber\\
\langle\hat{T}^{\phi}{}_{\phi}\rangle^{\rm s}_{\subHH}=
&
\frac{1}{64\pi^{2} r_{+}^{7}}\left\{\vphantom{\log\left(\frac{r_{+}-r_{-}}{r_{-}}\right)}r_{+}\left(r_{+}-2r_{-}\right)\left(5r_{+}+r_{-}\right)\right.\nonumber\\
&
\left.+2\left(r_{+}-r_{-}\right)\left[r_{+}^{2}\log\left(\frac{r_{+}-r_{-}}{r_{+}}\right)\right.\right.\nonumber\\
&
\left.\left.+2r_{-}^{2}\log\left(\frac{r_{+}-r_{-}}{r_{-}}\right)\right]\right\}+\mathcal{O}\left(r-r_{+}\right).
\end{align}
The components of the Polyakov RSET are of the same order of magnitude as the exact RSET and have positive sign at $r=r_{+}$ until $Q/M\approx0.94$. Despite both approximations being regular in the Hartle-Hawking state, the temporal and radial components of the $s$-wave RSET have opposite sign. This discrepancy is alarming, especially considering that the $s$-wave RSET only incorporates backscattering effects to the Polyakov RSET, and that both approximations should agree at the event horizon, where the gravitational potential vanishes. It has been argued that such a discrepancy is due to the dimensional reduction anomaly~\cite{NojiriOdintsov1999,Frolovetal1999,Sutton2000} that accounts for the non-commutativity of quantization and dimensional reduction. 

Among the various analytic approximations presented, only the Polyakov RSET can describe the Unruh state, for which diagonal RSET components are simply given by~\eqref{eq:UnruhPolyakov}. Hence, its components are finite at the horizon, contrary to the divergences appearing in the exact RSET. 

Turning now to the AHS-RSET, we have
\begin{align}\label{eq:RSETHorAHS}
\langle\hat{T}^{t}{}_{t}\rangle^{\rm AHS}_{\subHH}=
&
\langle\hat{T}^{r}{}_{r}\rangle^{\rm AHS}_{\subHH}=\frac{12 r_{+}^{2}-20r_{+}r_{-}+9r_{-}^{2}}{2880\pi^{2}r_{+}^{6}}\nonumber\\
&
-\frac{r_{+}^{2}-r_{-}^{2}}{64\pi^{2}r_{+}^{6}}\xi+\mathcal{O}\left(r-r_{+}\right),\nonumber\\
\langle\hat{T}^{\phi}{}_{\phi}\rangle^{\rm AHS}_{\subHH}=
&
\frac{33r_{+}^{2}-44r_{+}r_{-}+13r_{-}^{2}}{5760\pi^{2}r_{+}^{6}}\nonumber\\
&
-\frac{\left(r_{+}-r_{-}\right)^{2}}{32\pi^{2}r_{+}^{6}}\xi+\mathcal{O}\left(r-r_{+}\right).
\end{align}
The magnitude of the AHS-RSET components shows excellent agreement with the exact results. Similarly, we observe a sign inversion for certain coupling values. In the extremal limit $Q\to M$, the AHS-RSET becomes independent from $\xi$, in agreement with the exact results obtained in~\cite{Andersonetal1995}. Clearly, the AHS-RSET is perfectly regular at $r=r_{+}$ when the field is massless. For massive fields, it has an unphysical divergence at the horizon~\cite{AHS1995} caused by the failure of the WKB approximation there~\cite{Balbinotetal2001}.

In fact, for a minimally coupled massless field, the expression above for $\langle\hat{T}^{t}{}_{t}\rangle^{\rm AHS}_{\subHH}=\langle\hat{T}^{r}{}_{r}\rangle^{\rm AHS}_{\subHH}$ at $r=r_+$ is exact. To see this we first note that, using the results in \cite{BreenOttewill2012a} and Eq. (\ref{eq:RSET}), we have:
 \begin{align}
\langle\hat{T}^{r}{}_{r}\rangle^{}_{\subHH} |_{r=r_+}&=\langle\hat{T}^{r}{}_{r}\rangle^{}_{\subHH} |_{r=r_+}\\
&=- [g^{\phi \phi'}W_{,\phi\phi'}]_{r=r_+}+\frac{v_1(r_+)}{4\pi^2}. 
 \end{align}
 Moreover only the $n=0$ modes contribute to $[g^{\phi \phi'}W_{,\phi\phi'}]_{r=r_+}$ and for a massless field the radial mode functions $p_{0l},q_{0l}$ are given by Legendre polynomials. Hence it is straightforward to show using standard identities involving Legendre functions (see for example \cite{AHS1995}) that the $n=0$ mode sum contribution to $[g^{\phi \phi'}W_{,\phi\phi'}]_{r=r_+}$ vanishes, yielding:
\begin{align} 
\langle\hat{T}^{r}{}_{r}\rangle^{}_{\subHH} |_{r=r_+}=
 \frac{1}{4\pi^{2}}\left[v_1(r_+)-\mathcal{T}_{10}^{(p)}(r_+)-\mathcal{D}_{22}^{(-)}(r_+) \right]
\end{align}	
which for the Reissner-Nordström spacetime, evaluates to  $\langle\hat{T}^{t}{}_{t}\rangle^{\rm AHS}_{\subHH}=\langle\hat{T}^{r}{}_{r}\rangle^{\rm AHS}_{\subHH}$ with $\xi=0$. The $\langle\hat{T}^{\phi}{}_{\phi}\rangle_{\subHH}$ component on the event horizon contains a contribution from the $n=1$ radial modes and is therefore not amenable to a similar calculation, however we  note in passing that a quasi-closed form event horizon expressions for all RSET components for a scalar field in the Hartle-Hawking state with arbitrary mass and coupling, are presented in~\cite{BreenOttewill2012a}.

In summary, the Polyakov and AHS approximations are in good qualitative and quantitative agreement with exact results for massless fields in the Hartle-Hawking state. The $s$-wave approximation, on the contrary, predicts wrong signs for the energy density and radial pressure.

\subsubsection{The Boulware state}
To obtain the Polyakov, $s$-wave and AHS approximations in the Boulware vacuum, we need to subtract the corresponding temperature-dependent terms in each case, resulting in an RSET that is singular at the event horizon. 
For example, the $s$-wave RSET returns the following divergent contributions at $r=r_{+}$,
\begin{align}\label{eq:swaveBoul}
\langle\hat{T}^{t}{}_{t}\rangle^{\rm s}_{\subB}=
&
-\langle\hat{T}^{r}{}_{r}\rangle^{\rm s}_{\subB}=\frac{r_{+}-r_{-}}{384\pi^{2}(r-r_{+})r_{+}^{4}}+\mathcal{O}\left(r-r_{+}\right)^{0},\nonumber\\
\langle\hat{T}^{\phi}{}_{\phi}\rangle^{\rm s}_{\subB}=
&
-\frac{r_{+}-r_{-}}{64\pi^{2}(r-r_{+})r_{+}^{4}}\times\nonumber\\
&
\left\{1+\frac{2}{r_{+}}\log\left[\frac{\left(r-r_{+}\right)\left(r_{+}-r_{-}\right)}{r_{+}^{2}
}\right]\right\}\nonumber\\
&
+\mathcal{O}\left(r-r_{+}\right)^{0}.
\end{align}
For the analytic AHS-RSET we find
\begin{align}
\langle\hat{T}^{t}{}_{t}\rangle^{\rm AHS}_{\subB}=
&
-3\langle\hat{T}^{r}{}_{r}\rangle^{\rm AHS}_{\subB}=-\frac{\left(r_{+}-r_{-}\right)^{2}}{128\pi^{2}\left(r-r_{+}\right)^{2}r_{+}^{4}}\left(\xi-\frac{1}{6}\right)\nonumber\\
&
+\mathcal{O}\left(r-r_{+}\right)^{-1}\nonumber\\
\langle\hat{T}^{\phi}{}_{\phi}\rangle^{\rm AHS}_{\subB}=
&
-\frac{\left(r_{+}-r_{-}\right)^{2}}{192\pi^{2}\left(r-r_{+}\right)^{2}r_{+}^{4}}\left(\xi-\frac{19}{120}\right)\nonumber\\
&
+\mathcal{O}\left(r-r_{+}\right)^{-1}.
\end{align}
In the $\xi=0$ case, all RSET components in every analytic approximation diverge with the same signs at the event horizon, coinciding with the sign of the exact RSET for all $Q$ values. For all the coupling values analyzed, the sign of the AHS-RSET also agrees with that of the exact RSET.

The leading-order divergence in the $\langle\hat{T}^{t}{}_{t}\rangle^{\rm s}_{\subB}$ and $\langle\hat{T}^{r}{}_{r}\rangle^{\rm s}_{\subB}$ components~\eqref{eq:swaveBoul} comes from the Polyakov portion of the $s$-wave RSET. Corrections due to the gravitational potential are subleading in this state. However, these components diverge $\propto\left(r-r_{+}\right)^{-1}$, contrary to the AHS-RSET and the exact RSET that diverge $\propto\left(r-r_{+}\right)^{-2}$.
Here, we observe a major discrepancy between approximations, in the sense that temperature-dependent terms obtained from calculations in $2$D [see Eqs.~\eqref{eq:Polyakovtemp} and \eqref{eq:swavetemp}]
scale differently from those obtained in $4$D~\eqref{eq:AHStemp}. In the extremal limit, all three approximations are singular at the extremal horizon~\cite{Trivedi1992,Andersonetal1995}.

\subsubsection{Anomalous trace and massive fields}
Another quantity that is worth comparing is the trace of the RSET. The Polyakov RSET predicts the correct trace anomaly in $2$D. Due to the conformal symmetry of the dimensionally-reduced theory, the trace of the Polyakov RSET,
\begin{equation}
    \langle\hat{T}\rangle^{\rm P}=\frac{r\left(r_{+}+r_{-}\right)-3r_{+}r_{-}}{48\pi^{2} r^{6}},
\end{equation}
is state-independent in $4$D, thus finite and positive everywhere except when $Q$ is in the range \mbox{$M\geq |Q|>\sqrt{8/9}M$}. The $s$-wave RSET has a temperature-dependent trace instead obtained from the components~\eqref{eq:swaveRSET}. In the Boulware state, this trace is negatively divergent at $r=r_{+}$ for any $Q$, in stark contradiction with the trace of the exact RSET and the AHS approximation (see Eq.~\eqref{eq:AHStraceBoul} below). This divergence in the trace is caused by the angular pressures~\eqref{eq:swaveBoul}.

For the AHS-RSET the trace in the Hartle-Hawking state is
\begin{align}\label{eq:AHStrace}
    \langle\hat{T}\rangle^{\rm AHS}_{\subHH}=
    &
    \langle\hat{T}\rangle^{\rm AHS}_{\subB}+\frac{\kappa_+^{2}\left(\xi-\frac{1}{6}\right)} {16\pi^{2}r^{6}f^{2}}\times\nonumber\\
    &
   \left[2r_{+}^{2}r_{-}^{2}-2rr_{+}r_{-}\left(r_{+}+r_{-}\right)+r^{2}\left(r_{+}+r_{-}\right)^{2}\right]
\end{align}
where $\langle\hat{T}\rangle^{\rm AHS}_{\subB}$ is the zero-temperature part of the trace
\begin{align}\label{eq:AHStraceBoul}
    \langle\hat{T}\rangle^{\rm AHS}_{\subB}=
    &
    \frac{13r_{+}^{2}r_{-}^{2}-12rr_{+}r_{-}\left(r_{+}+r_{-}\right)+3r^{2}\left(r_{+}^{2}+r_{-}^{2}\right)}{720\pi^{2}r^{8}}\nonumber\\
    &
    -\frac{\left(r_{+}-r_{-}\right)^{2}\left(\xi-\frac{1}{6}\right)}{64\pi^{2}r^{6}f^{2}}\times\nonumber\\
    &
    \left[6r_{+}^{2}r_{-}^{2}-14rr_{+}r_{-}\left(r_{+}+r_{-}\right)\right.\nonumber\\
    &
    \left.+r^{2}\left(9r_{+}^{2}+32r_{+}r_{-}+9r_{-}\right)\right.\nonumber\\
    &
    \left.-20r^{3}\left(r_{+}+r_{-}\right)+12r^{4}\right].
\end{align}
In the conformally coupled case, Eq.~\eqref{eq:AHStrace} yields the correct trace anomaly. At the event horizon,~\eqref{eq:AHStraceBoul} is positively divergent for $\xi\leq1/6$ and any charge, whereas it is negatively divergent for $\xi>1/6$.

\begin{figure}
    \centering
    \includegraphics{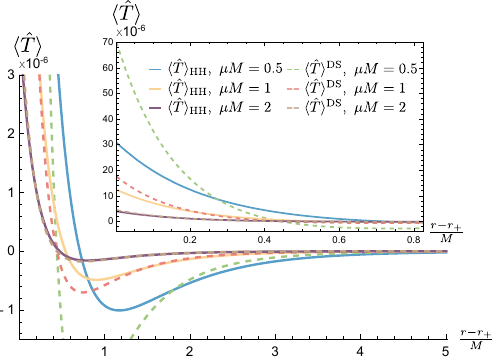}
    \caption{Trace of the RSET in the Hartle-Hawking state (continous lines) compared to the trace of the DS-RSET (dashed lines) for the cases $\mu M=\left\{1/2,1,2\right\}$ and $\xi=0$ from top to bottom, respectively. We have subtracted the Mc Laughlin tensor~\eqref{eq:McL} from the exact RSET to compare both results.}
    \label{fig:DS}
\end{figure}

Finally, Fig.~\ref{fig:DS} shows the trace of the exact RSET in the Hartle-Hawking state for fields of various masses, together with the trace of the DS-RSET. Clearly, the large-mass approximation becomes better as the field mass increases. However, this approximation does not include temperature-dependent terms and, as such, will not approximate the exact RSET evaluated in the Boulware and Unruh states near the event horizon. 

\section{Discussion and Conclusions}
In spherically symmetric black hole spacetimes, we can employ Euclidean methods to compute the RSET in the Hartle-Hawking state. The extended coordinate method outlined in this article provides an accurate and efficient way of calculating the RSET in such situations. On the other hand, the RSET for the Boulware and Unruh states cannot be defined on the Euclideanized metric. While progress is underway in applying the extended coordinate method to metrics with Lorentzian signature, we adopt an alternative approach towards obtaining the RSET in the Boulware and Unruh states by taking advantage of the fact that the difference between RSETs in two different Hadamard states is finite. Hence we need only ever renormalize in one reference state. In this paper, we used the Hartle-Hawking state as the reference state where renormalization is performed using the extended coordinate method on the Euclidean slice; the RSET in other states involves the components in the reference state plus finite, rapidly converging integrals over products of Lorentzian modes. This way, we were able to generate results for the RSET in the three states, for various couplings and field masses in the Reissner-Nordström spacetime. To the best of our knowledge, our results for massive fields in the Unruh state are the first in the literature.

The scope of this method extends to spacetimes in which there is no preferred thermal state, i.e., situations where the Euclideanized line element~\eqref{eq:metric} has no conical singularities. These encompass stellar spacetimes, where an auxiliary finite temperature state could be used as a conduit towards obtaining the RSET in the Boulware state. In stationary spacetimes that admit no Euclidean line element, as long as the RSET for a single state is known, it would still be possible to generate results for other states in those regions where both states are Hadamard.  

Regarding analytic RSET approximations, we have seen that the Polyakov approximation~\cite{ParentaniPiran1994} reproduces reasonably well the Hartle-Hawking and Unruh states, whereas it predicts a milder Boulware divegence at the horizon. The $s$-wave approximation~\cite{FabbriNavarro-Salas2005} is clearly off in the Hartle-Hawking state and is unable to yield a covariantly conserved RSET in the Unruh state. Note that both approximations only describe massless, minimally coupled fields. The AHS-RSET~\cite{AHS1995} admits any coupling but does not give standard results in Minkowski spacetime when the field is massive. For massless fields, it is qualitatively correct in the Hartle-Hawking and Boulware vacuums. Finally, massive fields in a state regular at the event horizon are well approximated by the DS-RSET. 

\appendix
\section*{Appendix A: DeWitt-Schwinger RSET}
Below you can find the coefficients in the components of the DS-RSET~\eqref{eq:DS-RSET},
\begin{align}\label{eq:DS-Coefs}
A_{0}=
&
-9308r_{+}^{3}r_{-}^{3},\quad A_{1}=14712\left(r_{+}+r_{-}\right)r_{+}r_{-},\nonumber\\
A_{2}=
&
-\left[6845\left(r_{+}+r_{-}\right)^{2}+7064r_{+}r_{-}\right]r_{+}r_{-},\nonumber\\
A_{3}=
&
\left[1237\left(r_{+}+r_{-}\right)^{2}+5137r_{+}r_{-}\right]\left(r_{+}+r_{-}\right),\nonumber\\
A_{4}=
&
-1125\left(r_{+}+r_{-}\right)^{2}-45r_{+}r_{-},\nonumber\\
B_{0}=
&
45864r_{+}^{3}r_{-}^{3},\quad B_{1}=-75936\left(r_{+}+r_{-}\right)r_{+}r_{-},\nonumber\\
B_{2}=
&
\left[36456\left(r_{+}+r_{-}\right)^{2}+77224r_{+}r_{-}\right]r_{+}r_{-},\nonumber\\
B_{3}=
&
-504\left[11\left(r_{+}+r_{-}\right)^{2}+67r_{+}r_{-}\right]\left(r_{+}+r_{-}\right),\nonumber\\
B_{4}=
&
5040\left(r_{+}+r_{-}\right)^{2},\nonumber\\
C_{0}=
&
1684r_{+}^{3}r_{-}^{3},\quad C_{1}=-3640\left(r_{+}+r_{-}\right)r_{+}r_{-},\nonumber\\
C_{2}=
&
\left[2081\left(r_{+}+r_{-}\right)^{2}+5145r_{+}r_{-}\right]r_{+}r_{-},\nonumber\\
C_{3}=
&
-\left[329\left(r_{+}+r_{-}\right)^{2}+2889r_{+}r_{-}\right]\left(r_{+}+r_{-}\right),\nonumber\\
C_{4}=
&
441\left(r_{+}+r_{-}\right)^{2}+657r_{+}r_{-},\nonumber\\
D_{0}=
&
-6552r_{+}^{3}r_{-}^{3},\quad D_{1}=-14112\left(r_{+}+r_{-}\right)r_{+}r_{-},\nonumber\\
D_{2}=
&
-\left[8232\left(r_{+}+r_{-}\right)^{2}+19880r_{+}r_{-}\right]r_{+}r_{-},\nonumber\\
D_{3}=
&
168\left[9\left(r_{+}+r_{-}\right)^{2}+65r_{+}r_{-}\right]\left(r_{+}+r_{-}\right),\nonumber\\
D_{4}=
&
-2016\left(r_{+}+r_{-}\right)^{2},\nonumber\\
E_{0}=
&
-13916r_{+}^{3}r_{-}^{3},\quad E_{1}=22808\left(r_{+}+r_{-}\right)r_{+}r_{-},\nonumber\\
E_{2}=
&
-\left[10947\left(r_{+}+r_{-}\right)^{2}+22771r_{+}r_{-}\right]r_{+}r_{-},\nonumber\\
E_{3}=
&
\left[1543\left(r_{+}+r_{-}\right)^{2}+10287r_{+}r_{-}\right]\left(r_{+}+r_{-}\right),\nonumber\\
E_{4}=
&
-1323\left(r_{+}+r_{-}\right)^{2}-1971r_{+}r_{-},\nonumber\\
F_{0}=
&
58968r_{+}^{3}r_{-}^{3},\quad F_{1}=-95424\left(r_{+}+r_{-}\right)r_{+}r_{-},\nonumber\\
F_{2}=
&
\left[45864\left(r_{+}+r_{-}\right)^{2}+92456r_{+}r_{-}\right]r_{+}r_{-},\nonumber\\
F_{3}=
&
-2352\left[3\left(r_{+}+r_{-}\right)^{2}+17r_{+}r_{-}\right]\left(r_{+}+r_{-}\right),\nonumber\\
F_{4}=
&
6048\left(r_{+}+r_{-}\right)^{2}.
\end{align}

\bibliographystyle{apsrev4-1}
	\bibliography{bib}
\end{document}